\documentclass[12pt,preprint]{aastex}
\usepackage{emulateapj5}
\newcommand{\etal}{{\it et al.\ }}

\newcommand{\eg}{{\it e.g.\ }}

\newcommand{\eion}{($\rm e^{-}\,+\,ion$)}
\newcommand{\lmlm}{\mbox{$\lambda\lambda$}}

\newcommand{\rne}{\mbox{$N_{\rm e}$}}
\newcommand{\kms}{\mbox{$\rm km\,s^{-1}$}}

\newcommand{\lya}{\mbox{$\rm Ly\,\alpha$}}
\newcommand{\lyb}{\mbox{$\rm Ly\,\beta$}}
\newcommand{\mum}{\mbox{$\rm \mu m$}}

\lefthead{Sigut \& Pradhan}
\righthead{\ion{Fe}{2} Emission from AGN}
 
\begin{document}

\submitted{Submitted to The Astrophysical Journal Supplement Series}
 
\title{Predicted \ion{Fe}{2} Emission-Line Strengths from Active Galactic Nuclei}

\author{T.\ A.\ A.\ Sigut}
\affil{Department of Physics and Astronomy,
The University of Western Ontario}
\authoraddr{London, Ontario Canada N6A 3K7}

\and

\author{Anil K. Pradhan}
\affil{Department of Astronomy, The Ohio State University}
\authoraddr{174 West 18th Avenue, Columbus, Ohio 43210-1106}

\begin{abstract}

We present theoretical \ion{Fe}{2} emission line strengths for physical
conditions typical of Active Galactic Nuclei with Broad-Line Regions.
The \ion{Fe}{2} line strengths were computed with a precise treatment
of radiative transfer using extensive and accurate atomic data from the
Iron Project.  Excitation mechanisms for the \ion{Fe}{2} emission
included continuum fluorescence, collisional excitation,
self-fluorescence amoung the \ion{Fe}{2} transitions, and fluorescent
excitation by \lya\ and \lyb. A large \ion{Fe}{2} atomic model
consisting of 827 fine structure levels (including states to $\rm
E\,\approx 15\;eV$) was used to predict fluxes for approximately 23,000
\ion{Fe}{2} transitions, covering most of the UV, optical, and IR
wavelengths of astrophysical interest.  Spectral synthesis for
wavelengths from $\lambda\,1600\,$\AA\ to $1.2\,\mum$ is presented.
Applications of present theoretical templates to the analysis of
observations are described. In particular, we discuss recent
observations of near-IR \ion{Fe}{2} lines in the
$8500\,$\AA--$1\,$\mum\ region which are predicted by the
\lya\ fluorescence mechanism. We also compare our UV spectral synthesis
with an empirical iron template for the prototypical, narrow-line
Seyfert galaxy I~Zw~1.  The theoretical \ion{Fe}{2} template presented
in this work should also applicable to a variety of objects with
\ion{Fe}{2} spectra formed under similar excitation conditions, such as
supernovae and symbiotic stars.

\end{abstract}


\section{Introduction}

Transitions of singly ionized iron dominate the spectra of many
astrophysical objects, from the sun and stars to active galactic nuclei
(AGNs) and quasars (Viotti 1988). However, the interpretation of this
spectrum is complex, and to extract meaningful results for the physical
conditions in the emitting region, and the iron abundance and
ionization fractions, one is faced with the solution of a complex
radiative transfer problem requiring the specification of many
thousands of radiative and collisional rates in a non-local
thermodynamic equilibrium (non-LTE) formalism.  Until recently, such
calculations have been hampered by the paucity of basic atomic data for
\ion{Fe}{2}.  However, the Iron Project (Hummer \etal\ 1993) has been
specifically initiated to address this problem, and new, accurate
atomic data for \ion{Fe}{2} have been calculated. In particular,
radiative dipole transition probabilities for over 21,000
fine-structure transitions of \ion{Fe}{2} have been computed by Nahar
(1995), and collision strengths for over 11,000 fine-structure
transitions have been computed by Zhang \& Pradhan (1995) and Bautista
\& Pradhan (1996). These calculations, and the Iron Project in general,
employ the powerful and accurate \underline{R}-matrix method (Burke \&
Berrington 1993).

In our earlier work on \ion{Fe}{2} (Sigut \& Pradhan 1998; SP98), we
employed a limited, non-LTE atomic model with 262 fine structure levels
which was still sufficiently large for \lya\ fluorescent excitation to
be investigated in detail. It was shown that \lya\ excitation can be of
fundamental importance in enhancing the UV and optical \ion{Fe}{2}
fluxes. In particular, it was predicted that \lya\ fluorescence results
in significant near-infrared \ion{Fe}{2} emission in the region
$\lambda\lambda\,8500-9500\,$\AA. Following the SP98 work, recent
observations have detected many of these near-IR \ion{Fe}{2} emission
lines from several narrow-line Seyfert~I galaxies (Rodriguez-Ardila
\etal 2001), and from a Type~IIn supernova remnant with narrow emission
lines (Fransson \etal 2001).  Although difficult to observe, these
near-IR \ion{Fe}{2} lines should be indicative of the excitation
mechanisms and the possible interplay between collisional and
fluorescent excitation (Rudy \etal 2000).  As our predicted near-IR
\ion{Fe}{2} fluxes in this wavelength region are likely to be of wider
interest, this paper presents a detailed line list from our non-LTE
calculations with an extended \ion{Fe}{2} model atom.

We intend this line list to be our first step in developing a reliable
set of {\it theoretical\/} templates for the iron emission from AGN.
Currently, due to the complexity of the observed iron emission from
AGN, such emission is typically modeled using empirical templates
derived from specific AGN spectra (Boroson \& Green 1992, Corbin \&
Boroson 1996). A recent example of this method is the
\ion{Fe}{2}-\ion{Fe}{3} template of Vestergaard \& Wilkes (2001)
derived from high-quality UV spectra of the narrow-line Seyfert~I
galaxy I~Zw~1.  Such templates play a critical role in extracting a
measure of the total iron emission from heavily blended and broadened
AGN spectra.  For example, Dietrich \etal (2002) apply the Vestergaard
\& Wilkes template to extract a measure the relative iron-to-magnesium
abundance ratio from a sample of high-$z$ quasars. Such studies seek to
constrain the epoch of major star formation in AGN using the
iron-to-magnesium abundance ratio as a nucleosynthesis ``clock"
following Hamann \& Ferland (1992). Such studies are beginning to
impose important cosmological constrains: for example, Aoki, Murayama
\& Denda (2002) have detected \ion{Fe}{2} emission from a $z=5.74$ QSO
with a strength comparable to much lower redshift objects.

Empirical templates have the advantage that they can side-step the
complicated process of specifying in detail the iron emission
mechanisms, and have generally found to provide better fits to
observations than theoretical templates (Iwamuro \etal\ 2002, Thompson,
Hill \& Elston 1999).  Nevertheless, there is still a strong need to
develop reliable theoretical templates:  (1) empirical templates assume
that the underlying AGN population used to construct the template is
typical and that the iron emission in other related objects can be
modeled as a simple scaling of the fiducial spectrum.  (2) Empirical
templates can never be completely free of the complications introduced
by the large blending and broadening present in AGN spectra. For
example, it is difficult to constrain the iron emission present at the
location of the \ion{Mg}{2} h \& k lines, although such an estimate
does affect the derived fluxes.

Theoretical iron flux templates can address both of these problems,
allowing estimates of the response of the iron emission to model
parameters which many differ from object to object (such as the
photoionizing radiation field), and providing a spectrum from which
complex blended features can be decomposed into their individual
contributions. But theoretical templates must first show that they can
explain the current empirical templates.  In this work, we will compare
our predictions with the empirical UV template of Vestergaard \&
Wilkes.

\subsection{The Physics of \protect\ion{Fe}{2} Line Formation in AGN}

The formation of the \ion{Fe}{2} emission spectrum from AGN is still
poorly understood (Joly 1993, Hamann \& Ferland 1999).  Typically,
photionization cloud models for the BLR fail to account for the
observed strength of the \ion{Fe}{2} emission. A class of
``super-strong'' \ion{Fe}{2} emitters is known (Lipari, Macchetto \&
Golombek 1991; Graham \etal\ 1996) which seem to be unaccountable by
traditional photoionized models.  In these cases, and possibly all, a
different cloud population may be the origin of the \ion{Fe}{2}
emission, such as mechanically heated clouds shielded from the central
continuum source (Joly 1987), perhaps originating in the outer regions
of an accretion disk (Collin-Souffrin \etal\ 1988).  An important
sub-class of AGN with BLRs are the narrow-line Seyfert galaxies
(Osterbrock \& Pogge 1985). I~Zw~1 is the prototypical narrow-line
quasar (Laor \etal 1997), and it is also a strong \ion{Fe}{2} emitter
(Marziani \etal 1996). Such narrow-line Seyfert~1 galaxies (NLS1)
enable both better emission line diagnostics and better tests of
theoretical spectra as their spectra are broadened with typical
velocities of $\leq 1000\,\rm km\,s^{-1}$.

The accepted micro-physics of \ion{Fe}{2} line formation in AGN is that
of Wills, Netzer \& Wills (1985) with the extension by Elitzur \&
Netzer (1985) to include fluorescent excitation by \lya. The
\lya\ excitation process is further studied by Johansson \& Jordan
(1984), Penston (1987), Sigut \& Pradhan (1998), and Verner \etal
(1999).  The proposed excitation mechanisms can be understood with the
aid of the highly simplified \ion{Fe}{2} energy level diagram shown in
Fig.~\ref{fig:fe2simp}, based on a similar figure by Penston (1987).
Four principal excitation mechanisms have been included:

\vspace{3mm}

(1) {\sc Continuum Fluorescence}: Photons incident on the
illuminated face of the BLR cloud are absorbed in the resonance
transitions and are subsequently re-radiated in the resonance and
optical lines. Strong optical emission, however, requires thermalization
of the resonance transitions in order to shift the effective branching
ratio towards optical emission. Thus this mechanism suffers from
having the photon source in the wrong location, namely outside the
cloud at small optical depth, something first noted by Netzer (1988).

(2) {\sc Collisional Excitation}: Inelastic collisions with electrons
excite the odd parity levels near 5~eV which then decay into the
optical and UV lines.  This mechanism is efficient whenever the gas
temperature is above $\approx 7000\,$K, temperatures which are
generally found in photoionized models of the BLR. Excitation is
irrespective of the local optical depth in the \ion{Fe}{2} lines, and
thus this mechanism does not suffer the limitations of continuum
fluorescence.  It is generally believed that collisional excitation is
responsible for the bulk of the \ion{Fe}{2} emission.

(3) {\sc Self-Fluorescence}:  Netzer \& Wills (1983) suggested that
self-fluorescence, that is absorption of the \ion{Fe}{2} UV resonance
photons by overlapping UV \ion{Fe}{2} transitions originating from the
odd parity levels near 5~eV (labeled ``unexpected UV" in
Fig.~\ref{fig:fe2simp}) was an important source of excitation to highly
excited states due to the large number of wavelength coincidences
between these groupings of levels.

(4) {\sc Fluorescent excitation by} \lya:  Penston (1987) noted that,
despite theoretical calculations to the contrary (Elitzur \& Netzer
1985), there is indirect evidence that \lya\ fluorescence may be an
important but overlooked excitation mechanism. Penston noted the
presence of unexpected UV \ion{Fe}{2} lines (see
Fig.~\ref{fig:fe2simp}) in the spectrum of the symbiotic star RR Tel
that seemed attributable only to cascades from higher levels pumped by
\lya\ fluorescence. The emission nebulae of symbiotics offer densities
and ionization parameters similar to those inferred for the BLRs of
AGN.  Graham \etal\ (1996) have identified emission from the UV
\ion{Fe}{2} multiplets expected to be preferentially strengthened by
this mechanism, as noted by Penston, in the spectrum of the
ultra-strong \ion{Fe}{2} emitter 2226-3905.

\vspace{3mm}

In this current work, in addition to considerably enlarging the
\ion{Fe}{2} atomic model to 827 fine structure levels up to $\rm
E\approx 15\,$eV, we make improvements to the modeling of all of
these excitation mechanisms:

(1) We solve the equation of radiative transfer with continuum
fluorescence included through the appropriate boundary conditions on
the transfer equations.

(2) We use a large and accurate set of {\underline R}-matrix
collision strengths for electron impact excitation of \ion{Fe}{2}. Such
rates are available for many of the key odd parity levels near 5~eV
with are the upper levels of most of the UV and optical emission.

(3) Line-overlap amoung the \ion{Fe}{2} transitions is included
exactly in the radiative transfer solutions. We considerably expand
the \ion{Fe}{2} atomic model to 827 fine-structure levels and over
23,000 radiative transitions.

(4) We include frequency-dependent source functions for \lya\ and
\lyb\ in the monochromatic source functions used in the radiative
transfer solutions.  While the \lya\ and \lyb\ source functions used
are approximate, as discussed in the Section~\ref{sec:calclya}, their
inclusion into the radiative transfer solution is exact.

\section{The Atomic Data}

The non-LTE calculations require a considerable amount of atomic data,
not only for \ion{Fe}{2} but also for several nearby ionization
stages.  \ion{Fe}{1} - \ion{Fe}{4} were considered in the calculation,
with \ion{Fe}{1}, \ion{Fe}{3}, and \ion{Fe}{4} represented as one level
atoms with ground state photoionization cross sections and total
radiative-plus-dielectronic recombination rates adopted from the
Ohio-State group calculations
(http://www.astronomy.ohio-state.edu/$\sim$pradhan; Bautista \& Pradhan
1995; Nahar \etal\ 1997; Nahar 1996; Nahar 1996; Bautista \& Pradhan
1997; Nahar \& Bautista 1999).  The \ion{Fe}{2} atomic model consisted
of 827 fine-structure levels representing the 265~$LS\pi$ multiplets
with observed energies (Johansson 1978).  An extensive \ion{Fe}{2} atom
is necessary for realistic estimates of the total emitted flux (Wills,
Netzer \& Wills 1985).  Rates between fine-structure levels, as opposed
to total $LS\pi$ terms, were used to make straightforward the inclusion
of line-overlap amoung the \ion{Fe}{2} transitions themselves and
fluorescent excitation by \lya\ and \lyb.  The doublet, quartet, and
sextet spin systems are illustrated in Figure~\ref{fig:fe2big} (as well
as the \lya\ fluorescence paths); a few octet terms ($\rm a^8S^e$, $\rm
z^8P^o$, $\rm a^8P^e$, and $\rm a^8D^e$) were included in the
calculation but are not shown in this Figure.

Radiative transition probabilities for \ion{Fe}{2} were drawn from
three sources: the critical compilation of Fuhr, Martin \& Wiese
(1988), the {\underline R}-matrix calculations for dipole-allowed
transitions ($\Delta S=0$) of Nahar (1995), and the extensive
semi-empirical calculations of Kurucz (1991, private communication). In
total, these sources give about 23,000 fine-structure transitions
between the included energy levels which satisfy $f_{ij}>10^{-4}$ and
$\lambda_{ij}<3\;\mu$m.

Photoionization cross sections for all \ion{Fe}{2} levels were adopted
from Nahar (1996), who computed them with the \underline{R}-matrix
method and also employed a unified treatment of radiative and
dielectronic recombination.  The full resonance structure of these
cross sections was retained. A state-specific
radiative-plus-dielectronic recombination coefficient was used for each
\ion{Fe}{2} quartet and sextet level. The remaining levels included
only radiative recombination through the Milne relation. Stimulated
recombination was also included, but generally makes a negligible
contribution to the total recombination rates.

For collisional excitation of \ion{Fe}{2} by electron impact, we adopt
the \underline{R}-matrix results of Zhang \& Pradhan (1995, denoted ZP)
and Bautista \& Pradhan (1996, denoted BP). The ZP calculation utilized
a 38-state close coupling target for \ion{Fe}{2}, mainly representing
the quartet and sextet symmetries of the $\rm 3d^6\,4s$, $\rm 3d^7$,
and $\rm 3d^6\,4p$ configurations. This calculation provides
collisional excitation rates to many of the key odd-parity levels
between 5 and 7~eV which are the upper levels of the resonance and
optical \ion{Fe}{2} emission lines.  The BP calculation used a somewhat
different target expansion designed to give collision strengths for
transitions amoung the low-lying even parity levels, including the
doublet system.  Collision strengths for electron excitation of all
remaining dipole-allowed transitions were computed with the effective
Gaunt factor approximation of Van~Regemorter (1969) with a Gaunt factor
of $0.3$.  Collisional coupling by electron excitation of the
fine-structure levels of each $LS\pi$ multiplet not explicitly computed
by ZP or BP were estimated by assuming that the highest energy set of
fine-structure collision strengths in each $LS\pi$ symmetry computed by
ZP or BP could be used as estimates for the fine-structure rates within
all higher $LS\pi$ multiplets.  If ZP or BP rates were unavailable for
any members in an $LS\pi$ symmetry, the fine-structure collision
strengths were set to 2.0 for $\Delta J=0,\pm 1$, and 0.5 otherwise.

Charge transfer recombination rates to \ion{Fe}{2} - \ion{Fe}{4} were
adopted from Kingdon \& Ferland (1996). Charge transfer ionization of
\ion{Fe}{2} was also included because recombination from \ion{Fe}{3} is
to the \ion{Fe}{2} ground state (Neufeld \& Dalgarno 1987).  All of
these charge transfer rates were computed in the Landau-Zener
approximation (see Flower 1990 for a discussion of this method), and
hence are not of the same accuracy as most of the atomic data used in
this work. We also note that the large rates predicted by this method
for the iron ions results in the domination of the charge-transfer
reactions in determining the iron ionization balance in AGN BLR
clouds.

\section{Calculations for \ion{Fe}{2} Fluxes} 

The coupled equations of radiative transfer and statistical equilibrium
were solved with the accelerated lambda-iteration method of Rybicki \&
Hummer (1991, 1992).  As overlap between the \ion{Fe}{2} transitions is
extensive and complex, the full preconditioning strategy suggested by
Rybicki \& Hummer (1992) was implemented. A diagonal approximate lambda
operator was used because of the large size of the iron atomic model.
Complete frequency redistribution (CRD) over depth-dependent Doppler
profiles was assumed for all of the \ion{Fe}{2} radiative transitions.
The width of the Doppler profile for each transition was assumed to be
\begin{equation} 
\Delta\nu_D = \frac{\nu}{c}\,\left(\frac{2kT_e}{m} +
\zeta_t^2\right)^{\frac{1}{2}} \,, 
\end{equation} 
where $\zeta_t$ is
the internal microturbulent velocity in the BLR clouds.

In order to solve the radiative transfer and statistical equilibrium
equations, it is necessary to know the structure of the BLR clouds,
that is the run of $T_e$ and $\rne$, the photoionizing radiation field,
and the background continuous opacities and source functions. To obtain
these quantities, we used the {\sc cloudy} code of Ferland (1991). The
cloud structure was fixed during the iron non-LTE calculation.  The
{\sc cloudy} models are traditional plane-parallel, one dimensional,
power-law illuminated clouds computed for a single ionization parameter
and total particle density, assuming constant total pressure. The shape
of the photoionizing continuum incident on the cloud was taken from
Mathews \& Ferland (1987). The particle conservation equation, which
closes the system of statistical equilibrium equations, was written at
each depth as
\begin{equation} 
\sum_{i=1}^{N} n_i = N_H \,
\epsilon_{Fe} \, \left( \sum_{j=1}^{4} f_j \right) , 
\end{equation}
where $N_H$ is the total hydrogen number density, $\epsilon_{Fe}$ is
the iron abundance relative to hydrogen, and $f_j$ is the fraction of
ionization stage $j$ of iron as predicted by {\sc cloudy}. The solar
iron abundance is $\log(N_{Fe}/N_H) = -4.52$.

\subsection{Fluorescent Excitation and Radiative Transfer}

\label{sec:calclya}

\lya\ pumping of the \ion{Fe}{2} emission was included by first
computing the frequency dependence of the \lya\ line emissivity. As
stimulated emission is not important, the emission profile for this
process can be set equal to the absorption profile without significant
error, and with this approximation, the frequency-dependent
\lya\ source function can be written as \begin{equation}
S^{Ly\alpha}_{\nu}=\left(\frac{n_{2p}\,A_{2p,1s}}{n_{1s}\,B_{1s,2p}-
n_{2p}\,B_{2p,1s}}\right)\;
\frac{\psi_{\nu}}{\phi_{\nu}}\;.  \label{eq:lyasnu} \end{equation} The
quantity in brackets is the frequency-independent source function valid
for complete redistribution. The absorption profile, $\phi_{\nu}$, was
taken to be a depth-dependent Voigt profile with a damping width set by
natural broadening. The emission profile, $\psi_{\nu}$, complicates the
problem considerably as it, in general, depends on the radiation field
(Mihalas 1978).  We have retained the {\sc cloudy} estimates of the
$1s$ and $2p$ \ion{H}{1} level populations in eq.~(\ref{eq:lyasnu}),
but have explicitly computed the emission profile using the
redistribution function \begin{equation} R(\nu',\nu) =
\gamma_c\,R_{II}(\nu',\nu) + (1-\gamma_c)\,\phi_{\nu'}\phi_{\nu}.
\end{equation} Here $\gamma_c$ is the fraction of coherently scattered
photons in the atom's rest frame.  Our treatment of the redistribution
function for resonance line emission, $R_{II}(\nu',\nu)$, follows
Venerazz, Avrett \& Loeser (1981) by using the partial coherent
scattering (PCS) approximation in a form suggested by Kneer (1975),
\begin{equation} R_{II}(\nu',\nu) \approx
\left<a\right>_{\nu}\,\phi_{\nu'}\,\delta(\nu'-\nu) +
(1-a_{\nu'\nu})\phi_{\nu'}\phi_{\nu}.  \end{equation} Here the function
$a_{\nu'\nu}$ effects the transition from complete redistribution in
the core to coherent scattering in the wings, and \begin{equation}
\left<a\right>_{\nu'\nu}=\int a_{\nu'\nu}\phi_{\nu'}\,d\nu'.
\end{equation} The relation between $a_{\nu'\nu}$ and
$\left<a\right>_{\nu}$ ensures the correct normalization of the
redistribution function.  We have used the form of $a_{\nu'\nu}$
described by Venerazz, Avrett \& Loeser (1981).

In the PCS approximation, the ratio of the emission to absorption
profile can be written as \begin{equation}
\frac{\psi_{\nu}}{\phi_{\nu}}=1+\gamma_c\,\left(\frac{\left<a\right>_{\nu}J_{\nu}-
\left<aJ\right>_{\nu}}{\bar{J}}\right), \end{equation} where
\begin{equation} \left<aJ\right>_{\nu} = \int
a_{\nu'\nu}\,\phi_{\nu'}J_{\nu'}\,d\nu', \end{equation} and
\begin{equation} \bar{J} = \int \phi_{\nu'}\,J_{\nu'}\,d\nu'.
\end{equation} The main limitation of the PCS approximation is that it
fails to account for Doppler diffusion in the coherent wings (Basri
1980). However, given the other approximations made in this work, the
use of the exact redistribution function seems unwarranted. We have
verified our treatment by matching the \lya\ profiles tabulated for BLR
clouds by Avrett \& Loeser (1988) using their models and hydrogen
populations (see Figure~\ref{fig:lyacheck}). We have tried both values
of $\gamma_c$ suggested by Avrett \& Loeser (1988), 0.998, the best
theoretical estimate, and 0.98, which gives the best match to the solar
profile. For all calculations, the \lya\ source function was calculated
independently and held fixed during the \ion{Fe}{2} solution.

\lyb\ was included in exactly the same manner as \lya, except that the
fraction of coherent scattering was assumed to be 0.4. The $n=1$ and 3
level populations were also adopted from the {\sc cloudy} model.

In the radiative transfer solutions, fluorescent excitation by
\ion{Fe}{2} line overlap and by \lya\ (or \lyb with analogous expressions) 
was included by constructing the
total monochromatic source function at each frequency as
\begin{equation}
S_{\nu}=\sum_{l}\,\frac{\chi_{\nu}^{l}}{\chi_{\nu}}\,S^{l}+
\frac{\chi_{\nu}^{c}}{\chi_{\nu}}\,S^{bck} +
\frac{\chi^{Ly\alpha}_{\nu}}{\chi_{\nu}}\, S^{Ly\alpha}_{\nu},
\end{equation} where $S^{l}$ is the CRD line source function for each
contributing iron transition, $S^{bck}$ is the background continuum
source function, and $S_{\nu}^{Ly\alpha}$ is the
\lya\ frequency-dependent PRD source function.  The total monochromatic
opacity at frequency $\nu$ is given by \begin{equation}
\chi_{\nu}=\sum_{l}\,\chi_{\nu}^{l}+\chi_{\nu}^{c}+\chi_{\nu}^{Ly\alpha}.
\end{equation} where $\chi_{\nu}^{l}$ is the opacity due to each
contributing iron transition, $\chi_{\nu}^{c}$ is the background
continuous opacity, and $\chi_{\nu}^{Ly\alpha}$ is the opacity due
\lya. The \lya\ source function and opacity were not assumed to be
constant across the iron line profiles.  Table~\ref{tab:lyalines} lists
the \ion{Fe}{2} transitions originating from $a\,^4\!D^e$ within $\pm
3\;$\AA\ of \lya. Critical parameters in the strength of this pumping
are the oscillator strengths of the \ion{Fe}{2} transitions. As shown
in the Table, the recent results of Nahar (1996) and Kurucz (1991,
private communication) generally estimate oscillator strengths about an
order of magnitude larger than Kurucz (1981).

The total line flux in each \ion{Fe}{2} transition was computed with
two methods:  First, as frequency-angle dependent radiative transfer is
explicitly solved, the fluxes can be computed from the emergent line
profiles, relative to the continuum, just as in the analysis of
observations, \begin{equation} F_{ji}=\int_{\Delta\nu_{ij}}
(F^{i}_{\nu}-F^{i}_{c})\,d\nu +
       \int_{\Delta\nu_{ij}} (F^{s}_{\nu}-F^{s}_{c})\,d\nu \,.
\label{eq:flx1} \end{equation} Here, $\Delta\nu_{ij}$ is the
integration bandwidth for the $i-j$ transition, and $F_{c}$ is the
continuum flux at the line's frequency.  The first term is the emission
from the illuminated face, and the second, the emission from the
shielded face.

Another way of computing the emergent flux is from the cooling
function. Considering {\em only} the line $i-j$, the first moment of
the transfer equation integrated over frequency gives \begin{equation}
\frac{dF^{o}_{ji}}{dz}=4\pi\;\int_{\Delta\nu}\,\chi_{\nu}^{l}\,(S^{l}-J_{\nu})
\; d\nu.  \end{equation} This clearly represents the cooling (or
heating) due to the $i-j$ transition in $\rm ergs\,cm^{-3}\,s^{-1}$.
Substituting for the line opacity and the line source function in terms
of the level populations, and defining the {\em net radiative bracket}
for the transition $i-j$ as \begin{equation} \rho_{ij}\equiv
1-\frac{\bar{J}_{ij}}{S^{l}} \,, \label{eq:nrb} \end{equation} we have
\begin{equation}
\frac{dF^{o}_{ji}}{dz}=h\nu\,A_{ji}n_{j}\,\rho_{ij}\equiv \Phi_{ij}(z)
\,.  \label{eq:cool} \end{equation} Thus we can define a net line flux
escaping from both sides of the cloud as the integral of the cooling
function over depth \begin{equation} F^{o}_{ij}\equiv \int\,
\Phi_{ij}(z)\,dz \,.  \end{equation} This is a very convenient
expression for the flux and provides the most direction connection
between the flux predicted by exact radiative transfer and traditional escape
probability methods. In the later case, an escape probability function
replaces the net radiative bracket.  However, it is important to
realize that  $F^{o}_{ji}\neq F_{ji}$, as the latter is defined as the
flux relative to the continuum, in the absence of the line. Performing
a similar analysis for the continuum alone, it is trivial to show that
the relation between these two fluxes is 
\begin{equation}
F_{ji}=F^{o}_{ij}+4\pi\int_{\Delta\nu_{ij}} (\chi_{\nu}^{c}-\sigma_{\nu})\;
(J_{\nu}^{c}-J_{\nu})\;d\nu.  \label{eq:flx2} \end{equation} 
Here,
$J_{\nu}^{c}$ is the mean intensity in the continuum in the absence of
the line.

Generally, both equations~(\ref{eq:flx1}) and (\ref{eq:flx2}) have been
used to compute the emergent fluxes in the lines. As these two methods
compute the fluxes in very different ways, agreement between the two
methods is a good check on the consistency of the calculation (Avrett
\& Loeser 1988).  Agreement in the current work was generally within
few percent.

\section{Results and Discussion}

\ion{Fe}{2} spectra were computed for four BLR cloud models typical of
the conditions thought to exist in the \ion{Fe}{2} emitting clouds.
The calculations have been made for traditional clouds of a single
specified density and ionization parameter, as opposed to the more
realistic {\it locally optimally emitting} cloud models of Baldwin
\etal (1995), as the main interest of the current work is the interplay
of the various iron emission excitation mechanisms and not the detailed
structure of the BLR.  Table~\ref{tab:agn} lists the basic BLR
parameters, along with the total H$\beta$ flux predicted by {\sc
cloudy} which is used to normalize the total \ion{Fe}{2} flux of
Figure~\ref{fig:fe2cog}.  For each basic cloud model, several values of
the cloud's internal turbulent velocity ($0$, $10$, $20$ and
$40\;$\kms) and iron abundance ($1/3$ solar, solar, and $3$ times
solar) were adopted.  Detailed tabular fluxes and spectral-synthesis
plots are presented for one of these models, model A, for a calculation
including both \lya\ and \lyb\ fluorescent excitation. Predicted fluxes
for the 600 strongest \ion{Fe}{2} transitions are given in Table~3, and
Figures~\ref{fig:flx_1600_2300} to \ref{fig:flx_7000_12000} show the
predicted \ion{Fe}{2} spectrum.  Two spectra are shown, one broadened
by convolution with a dispersion profile with a width of 100\,\kms, and
one broadened by 500\,\kms; these atypically low broadenings are
selected for clarity, with 500~\kms\ approaching the lowest observed
values for AGN (\eg 1~Zw~I) and 100~\kms\ appropriate for the
"nano-quasars'' (accreting white dwarfs) of Zamanov \& Marziani
(2002).  The BLR clouds were assumed to have a covering fraction of 5\%
of the central continuum source with equal contributions from the
illuminated and shielded cloud faces.

The predicted \ion{Fe}{2} fluxes from these models generally cover most
of ultraviolet (UV), optical (Opt), and infrared (IR) wavelengths of
interest in astrophysical sources.  While the line fluxes from the
present non-LTE models are particular to the assumed BLR conditions of
AGNs, similar excitation mechanisms and conditions may prevail in other
sources, such as symbiotic stars and supernova. In such cases, the
\ion{Fe}{2} line list of Table~3 should be useful in the identification
of observed features or the absence thereof. The line fluxes given in
Table~3 utilize a model atom which is considerably larger
than previous works.  For example, Verner \etal (1999) consider 371
levels, up to 11.59 eV, partially using the Iron Project data described
in section 2. Also their radiative transfer treatment is approximate,
employing escape probability methods, which, while being easier, may
not be sufficiently precise for an accurate treatment of the strong
\lya\ excitation of \ion{Fe}{2}.

The dependence of the total predicted UV+Opt \ion{Fe}{2} flux, relative
to H$\beta$, on the internal microturbulence ($\zeta_t$), the total
iron abundance ($\epsilon_{\rm Fe}$), and the inclusion or exclusion of
\lya and \lyb fluorescent excitation, is shown in
Figure~\ref{fig:fe2cog} for each of the models of Table~\ref{tab:agn}.
When looking at the trend with abundance, it should be kept in mind
that the predicted net \ion{Fe}{2} cooling has not been used to
recompute the {\sc cloudy} BLR model; thus, the trends are likely
overestimates of the real effect because of the strong ``thermostatic''
effect of the \ion{Fe}{2} cooling.  These figures clearly show the
importance of \lya\ fluorescent excitation on the \ion{Fe}{2} fluxes.
The influence of \lya\ is largest for the higher ionization parameter,
$U_{ion}=10^{-2}$, and for the lower particle density, $\log(N_H)=9.6$;
the higher particle density tends to thermalize the \lya source
function which reduces the net pumping rate.  The total predicted
\ion{Fe}{2} flux, relative to H$\beta$, is largest for the lower
ionization parameter models, $U_{ion}=10^{-3}$, and the lowest density,
$\log(N_H)=9.6$. The assumed internal cloud microturbulent velocity, a
poorly known parameter, is also seen to have a large impact on the
predicted fluxes at higher abundances. To extract realistic abundances
for the BLR, a clear procedure needs to be developed to constrain this
parameter.

The distribution of \ion{Fe}{2} fluxes in model~A with and without
\lya\ fluorescent excitation is illustrated in
Figure~\ref{fig:flx2cum}.  The upper-left panel shows the cumulative
predicted flux, including \lya and \lyb pumping, as the \ion{Fe}{2}
fluxes are summed from the strongest to weakest transition.  The 50\%
point occurs after only the 120 strongest transitions, whereas the 90\%
point is reached after 1250 transitions.  The figure also shows that
the cumulative flux distribution is reasonably well fit by a function
of the form \begin{equation} F(n)=\sum_{i=1}^{n}\,F_i=F_{\rm
tot}\left(1-e^{-(n/N_o)^{\alpha}}\right) \label{eq:cumm} \end{equation}
with $\alpha\approx 0.53$ and $N_o\approx 240$. Here $F_{\rm tot}$ is
the total \ion{Fe}{2} flux. The bottom-left panel shows a histogram of
the actual distribution in the $\log_{10}$ of the \ion{Fe}{2} fluxes;
the distribution is seen to be double-peaked which is a consequence of
\lya\ fluorescent excitation.  The rightmost panels of
Figure~\ref{fig:flx2cum} illustrate the case excluding
\lya\ fluorescent excitation. In this case, roughly half the number of
transitions are required to carry 50\%, and 90\% of the total flux,
namely 50 and 625 transitions respectively, and the flux distribution
is no longer double-peaked.  The best-fit parameters of
Equation~\ref{eq:cumm} for the cumulative flux distribution are
$\alpha\approx 0.48$ and $N_o\approx 107$.

An alternative view of the \ion{Fe}{2} flux distribution is given in
Figure~\ref{fig:flx2hist}. This Figure shows a histogram of the predicted
\ion{Fe}{2} flux, including \lya and \lyb pumping, based on the
energy of the upper level of each transition. Again, two peaks in the
distribution can be seen, one near $0.4-0.5\,$Ryd where the flux
originates from lines collisionally excited from the ground
configurations to the low-lying, odd-parity levels, and a second one at
higher energies corresponding to the initial cascades from
\lya\ pumping. The lower panel shows a histogram of the number of
\ion{Fe}{2} energy levels as a function of energy. Again, BLR model~A
was employed for these calculations.

\subsection{Theoretical Templates}

The predicted line fluxes, such as those of Table~3 may serve as the
basis for a set of theoretical templates for the \ion{Fe}{2} emission
from AGN covering the entire wavelength region from the UV to the
near-IR.  Subsets of these templates may be used to study individual
\ion{Fe}{2} features or to subtract the total contribution of
\ion{Fe}{2} in a particular range. Sample subsets of the UV line fluxes
in the $1600-3100\,$\AA\ region are shown in
Figures~\ref{fig:flx_1600_2300}, \ref{fig:flx_2300_2700}, and
\ref{fig:flx_2700_3100}, a region which contains amoung the strongest
\ion{Fe}{2} emission features.

Figure~\ref{fig:1zw1_uv} compares our predicted UV fluxes with the
empirical \ion{Fe}{2}-\ion{Fe}{3} template of Vestergaard \& Wilkes
(2001). The underlying cloud model used for the calculation is model~B
of Table~\ref{tab:agn}, which was found to give the best fit amoung the
four models. The calculated spectrum was convolved with a dispersion
profile of a width of 900\,\kms and then normalized to the average
relative flux of the observed template spectrum in the range of
$\lambda\,2400$-$2500\,$\AA. The figure shows the computed spectrum
both with, and without, fluorescent excitation for several combinations
of the internal cloud microturbulence ($\zeta_t$) and iron abundance
($\epsilon_{\rm Fe}$).  A strong constraint on the model fits is the
flatness of the observed UV emission between $2300$ and $2600\,$\AA.
Overall, the observed spectrum can be reproduced in broad outline,
although many individual features remain discrepant. However, we are
not trying to propose a specific BLR model for I~Zw~1, but simply to
compare its \ion{Fe}{2} spectrum to the current predictions of our BLR
models.  Another UV comparison with I~Zw~1 is shown in
Fig.~\ref{fig:1zw1}. Here, the computed \ion{Fe}{2} template in the
interesting $1500-2200\,$\AA\ wavelength range is compared to
observations of Marziani \etal 1996. Several of the UV features (albeit
blended) correspond well to the observed ones, thus facilitating
identification and determination of lines from \ion{Fe}{2} and other
ionic species.

Our predicted optical and IR line fluxes are shown in
Figures~\ref{fig:flx_3100_7000} and \ref{fig:flx_7000_12000}, covering
wavelengths from $3100\;$\AA\ to $1.2\,$\mum.  These lines may be
resolved through high-resolution spectroscopy and can provide useful
plasma diagnostics of the emitting regions.  The optical/IR lines can
be divided into two groups: those from amoung the low-lying levels of
\ion{Fe}{2} which depend primarily on the electron temperature and
density, and those from amoung the high-lying levels which are populated
primarily by cascades from upper levels and are therefore dependent on
fluorescent excitation and ($\rm e^{-}$+\ion{Fe}{3}) recombination.
Because the atomic physics considerations are different, one may
distinguish between the two groups through correlations between the
observed and theoretical spectra, as discussed in the next section.

\subsection{Fluorescent Excitation of Optical and Infrared Lines}

The ground term of \ion{Fe}{2} is the high multiplicity sextet
$3d^6\,4s\,^6\!D_{\frac{9}{2},\frac{7}{2},\frac{5}{2},\frac{3}{2},\frac{1}{2}}$.
The transitions potentially affected by ultraviolet continuum
fluorescence from the ground state should therefore be other sextets of
opposite parity, \eg\ $3d^6\,4p\,^6\!(D,F,P)^o_J$, all around 6~eV,
resulting in transitions usually labeled {\it uv1, uv2, uv3}
respectively. These can influence the low-lying optical and IR
transitions via cascades, as shown in Figure~\ref{fig:fe2_lowe}, but
two effects diminish the importance of continuum fluorescence: (1) the
number of excited sextet levels is relatively small and the subsequent
cascades are via inter-combination transitions
(Figure~\ref{fig:fe2_lowe}), and (2) the source is outside the BLR
cloud and photons cannot penetrate to layers where the effective
branching ratio favors optical-infrared emission.  Therefore
ultraviolet pumping from the ground state is not likely to be a
dominant mechanism for the enhancement of line emissivities of optical
and IR lines amoung low-lying levels, although individual line ratios
may be affected (Bautista \& Pradhan 1998). Excitation from the
ground $a\,^6\!D$ state, at temperatures characteristic of \ion{Fe}{2}
emitting region of $\approx\,10^{4}$ K, proceeds mainly through
electron impact excitation and produces only the low-lying
optical-infrared lines shown in Figure~\ref{fig:fe2_lowe}.  Lines
originating from higher levels must be excited through other
mechanisms, either through ($\rm e^{-}$~+~ion) recombination or
fluorescent excitation from low-lying metastable levels which are
collisionally populated.

As most of the excited levels of \ion{Fe}{2} are of quartet
multiplicity, fluorescent excitation is most likely to involve
transitions between quartet symmetries originating from low-lying, even
parity, metastable levels. For \lya\ pumping in particular, the most
likely transitions are from the lowest excited quartet state, $3d^6
\,4s\;a\,^4\!D^e$ to odd parity excited quartet levels (Johansson \&
Jordan 1984).  Table~\ref{tab:lyalines}, Figure~\ref{fig:fe2big}, and
Figure~\ref{fig:atom_fe2_lyair}
show such levels accessible within $\pm\,3\;$\AA\ of \lya. As shown in
SP98, the initial decays from \lya\ excitation of $a\,^4\!D^e$ results
in a group of lines in the $8000-9500\;$\AA\ range, with a large
predicted feature at about $9200\;$\AA. This excitation decay path is
explicitly shown in Figure~\ref{fig:atom_fe2_lyair}.

These sub-\mum\ lines have recently been detected in a supernova
remnant, SN~1995N, which is a Type IIn supernova with narrow emission
lines originating from circumstellar gas photoionized by X-rays from
the shock (Fransson \etal\ 2001). Although the conditions do not quite
correspond to the BLRs of AGN, there is still striking agreement
between the observed spectra from $8000-9500\;$\AA\ and the theoretical
fits including the \ion{Fe}{2} line fluxes computed by the SP98 models.
In addition, Fransson \etal\ find that the corresponding \ion{Fe}{2}
lines, $\lmlm 2506/2508\;$\AA, indicative of this \lya\ pumping path
(see Figure~\ref{fig:atom_fe2_lyair}), are the strongest such features
in the UV. We note that in our comparison to the UV spectrum of I~Zw~1
(Figure~\ref{fig:1zw1_uv}) the \ion{Fe}{2} feature just redward of
\ion{Mg}{2} h \& k is only reproducible by our models which include
\lya\ fluorescent excitation.

\lya\ fluorescence can also involve excitation from several other
low-lying, metastable levels (see Figure~\ref{fig:fe2big}). An
important example is excitation from the $a\,^4\!G^e$ levels about 3~eV
above the ground state.  The resultant cascades can form a series of
\ion{Fe}{2} lines near 1~\mum.  Figure~\ref{fig:atom_fe2_lyair} also
depicts this excitation/decay path in quartet multiplet.

These 1~\mum\ lines have been discussed by Rudy \etal\ (2000) in
observations of I~Zw~1, and by Rodriguez-Ardila \etal\ (2001) in a
sample of narrow-line Seyfert galaxies. Rodriguez-Ardila \etal\ also
provide the first AGN observations of the primary cascade lines of
\ion{Fe}{2} near $\lambda\,9200\;$\AA\ due to \lya  pumping.  As noted,
the excitation-cascade mechanism for the 1~\mum\ lines entails
excitation via $a\,^4\!G^e \rightarrow (t,u)\,^4\!G^o$, followed by
downward UV transitions to $b\,^4\!G^e$ via $\lmlm\,1870/1873$ and
$1841/1845\;$\AA\ (see Figure~\ref{fig:atom_fe2_lyair}). Further
transitions within the $b\,^4\!G^e-z\,^4\!F^o$ multiplet then gives
rise to the lines near 1~\mum. Rodriguez-Ardila \etal also report
these lines from narrow-line Seyferts similar to I~Zw~1. Earlier, Rudy
\etal\ argued against the \lya\ fluorescent mechanism on the grounds
that the UV feed lines are not seen in the spectrum of I~Zw~1 (see Laor
\etal\ 1997), thus suggesting electron collisional excitation as a
viable mechanism.  In their analysis, however, Rudy \etal\ did not
consider one of the UV multiplets, $a\,^4\!G^e \rightarrow u\,^4\!G^o$,
that give rise to the $1841$/$1845\,$\AA\ lines.  Our inclusion of both
the $(t,u)\,^4\!G^o$ levels is based on theoretical calculations of
\ion{Fe}{2} energy levels and transition probabilities of Nahar
(1995).  The observations of Rodriquez-Ardila \etal\ (2001) confirm
that \lya\ pumping is a strong contributor to the formation of 1~\mum
lines (suggested in this work and SP98), although collisional
excitation, and possibly \eion\ recombination, may be needed to account
for the observed intensities (see Figure~\ref{fig:flx_7000_12000}).
Discrepancies with observations may also be attributable to
uncertainties in our atomic data, particularly the collision
strengths.

\section{Conclusions}

Multi-level, accelerated lambda-operator techniques for non-LTE
radiative transfer now allow solutions to be obtained for highly
realistic atomic models including complex cases of line-overlap and
fluorescent excitation. Coupled with the new atomic data from the Iron Project,
such techniques have been applied for the first time with a reasonably
complete \ion{Fe}{2} atomic model to theoretical AGN BLR spectra. The
theoretical \ion{Fe}{2} line fluxes presented should help in the
identification of \ion{Fe}{2} transitions in AGNs and related sources,
and in the delineation of excitation mechanisms producing the \ion{Fe}{2}
spectrum.

We are extending the calculations to include the line spectra of other
iron ionization stages, principally Fe\,{\sc i} and Fe\,{\sc iii}.
Laor \etal\ (1997) and Vestergaard \& Wilkes (2001) specifically note
the present of significant \ion{Fe}{3} in the spectrum of I~Zw~1, and
Graham \etal\ (1996) have detected Fe\,{\sc iii} emission in the
ultra-strong Fe\,{\sc ii} emitter 2226-3905.  Kwan \etal\ (1995) have
detected Fe\,{\sc i} emission in two Fe\,{\sc ii}-strong quasars, IRAS
07598+6508 and PHL 1092.  Simultaneous modeling of these ionization
stages should provide more constraints on the nature of the iron
emission.

Our calculations currently include transitions between observed energy
levels whereas there remains a large number of theoretically predicted
energy levels.  We are including these levels and the implied radiative
transitions in order to provide a much more complete description of the
Fe\,{\sc ii} emission spectrum.

We are currently working on bringing the entire photoionization
calculation for all atoms and ions within the framework of exact
radiative transfer established in this work.  This will allow a
self-consistent treatment of the Fe\,{\sc ii} emission by including it
in the net heating/cooling which determines the temperature structure,
and will also allow an {\em exact\/} treatment of \lya\ fluorescent
excitation of Fe\,{\sc ii} emission.

We are also working on several other specific problems, such as the
interpretation of Fe\,{\sc ii}/Mg\,{\sc ii} line ratios, employing
extended non-LTE models for the relevant atomic species and exact
radiative transfer.

The tabular and graphical material presented in this work is available
electronically on request.

\vspace{0.3in}

\noindent
We would like to thank Sultana Nahar for numerous contributions, and
Marianne Vestergaard for the data in Figure~\ref{fig:1zw1_uv}.  This work was supported
by the Natural Sciences and Engineering Research Council of Canada
(TAAS), and by the U.S. National Science Foundation and NASA (AKP).

%
% Figure captions
%

\newpage

\setcounter{figure}{0}

\begin{figure}[p]
\epsscale{0.8}
\plotone{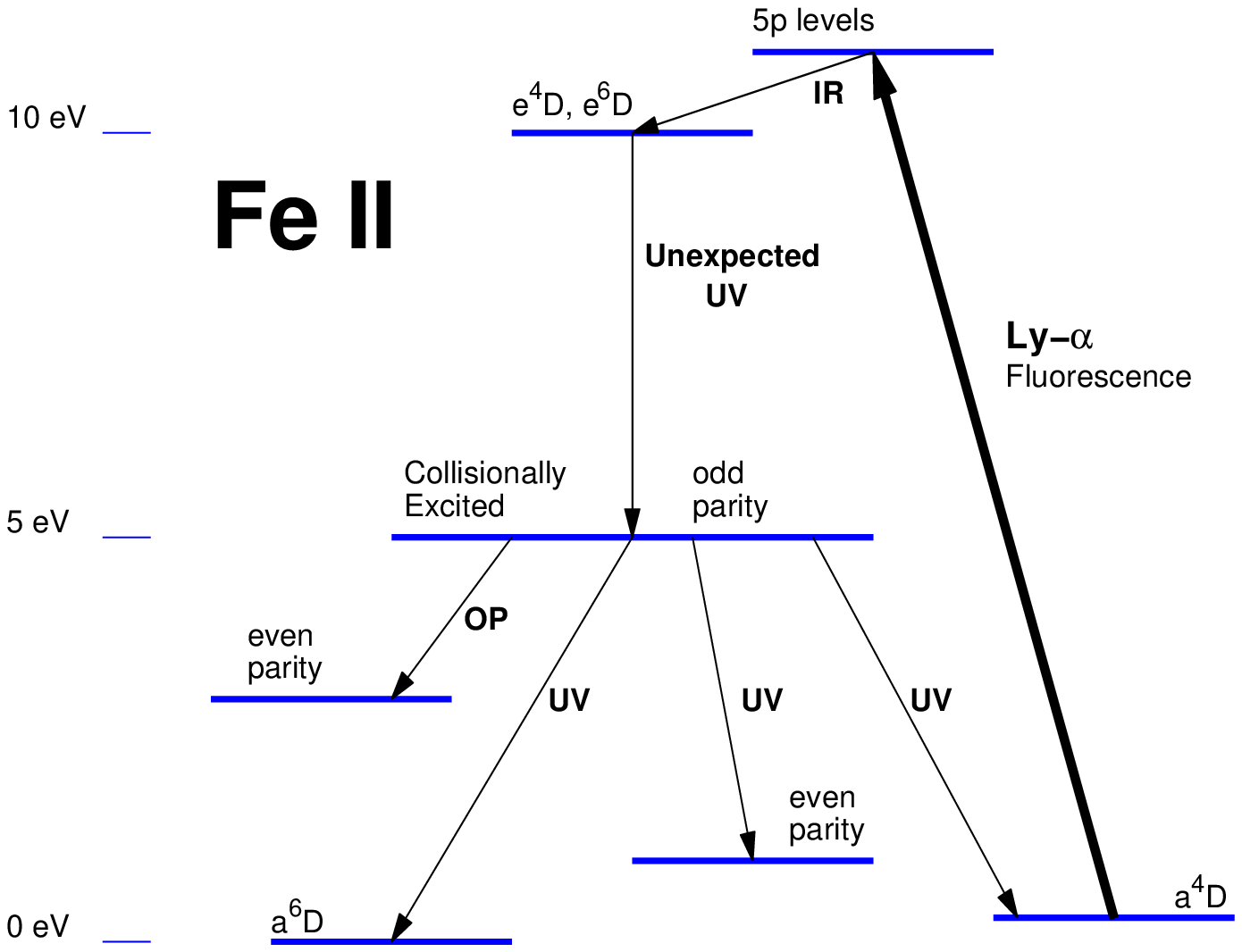}
\figcaption[f1.eps]{A highly simplified \protect\ion{Fe}{2}
Grotrian diagram illustrating the main excitation mechanisms discussed
in the text. \label{fig:fe2simp}}
\end{figure}

\begin{figure}[p]
\epsscale{1.0}
\plotone{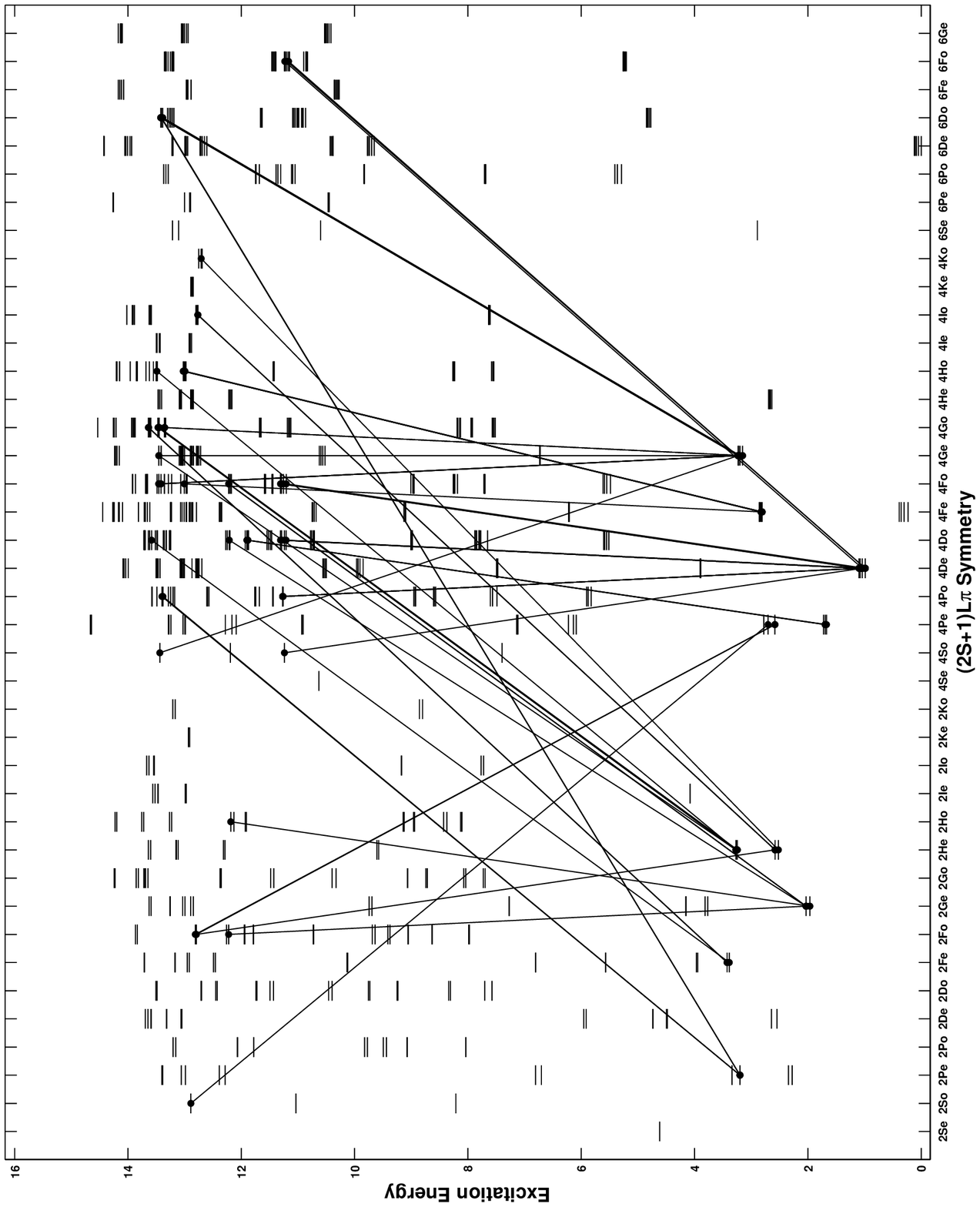}
\figcaption[f2.eps]{An Grotrian diagram showing all 827 \ion{Fe}{2} energy
levels (with the exception of a few levels in the octet spin system).
Solid lines connect transitions within $\pm\,3\,$\AA\ of
\protect\lya\ and represents only a small fraction of the $23,000$
radiative transitions included in the calculations.
\label{fig:fe2big}}
\end{figure}

\begin{figure}[p]
\plotone{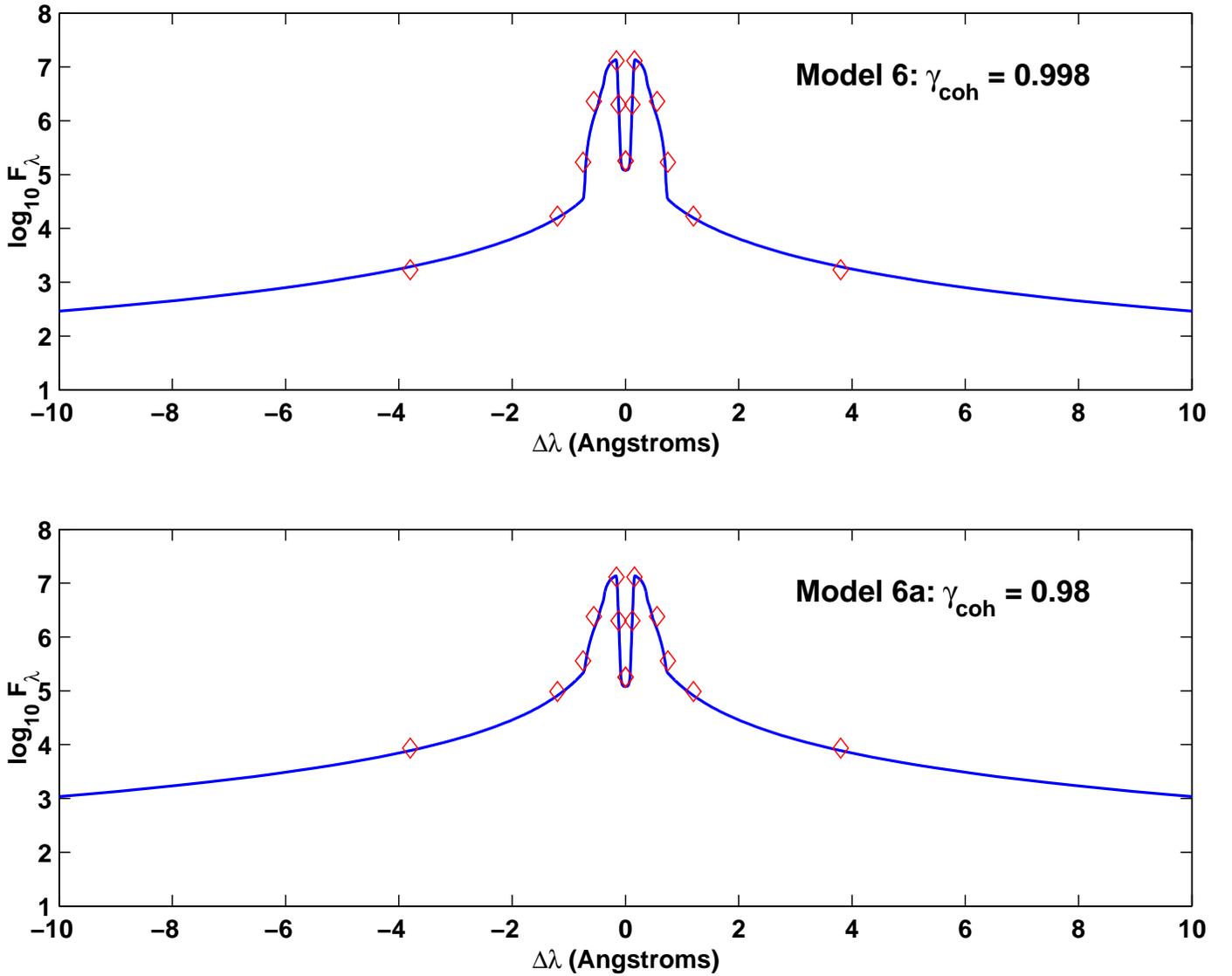}
\figcaption[f3.eps]{Comparison between the \protect\lya\ profiles of Avrett \&
Loeser (1988, diamonds) with those calculated in the current work
using the same BLR model (solid lines). The fraction of coherently
scattered \protect\lya\ photons, $\gamma_{\rm coh}$, is indicated in
each panel.  \label{fig:lyacheck}}
\end{figure}

\begin{figure}[p]
\plotone{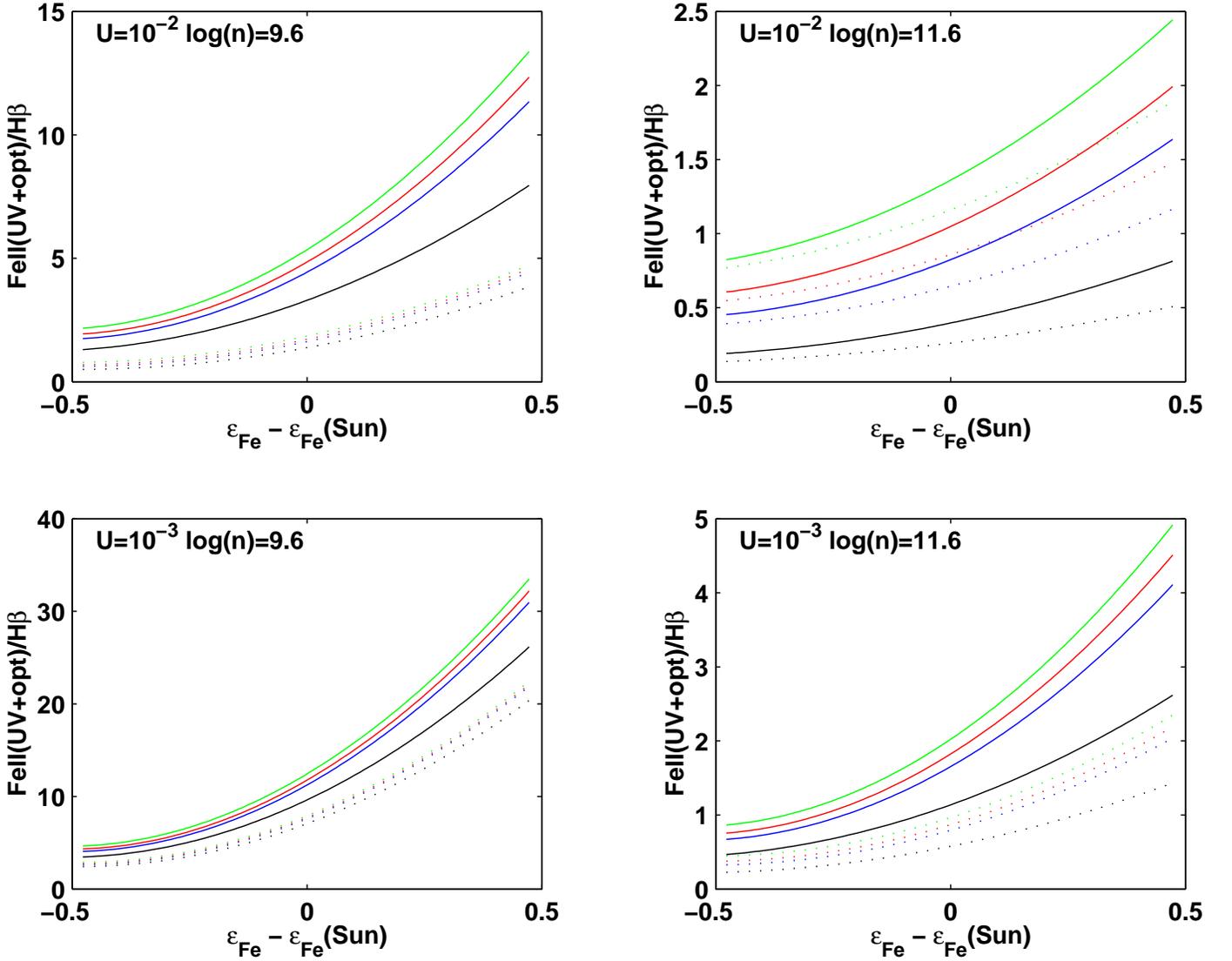}
\figcaption[f4.eps]{The dependence of the total predicted \protect\ion{Fe}{2}
UV+Opt flux on the iron abundance ($\epsilon_{Fe}$) and internal cloud
microturbulent velocity ($\zeta_t$). Each panel represents one of the
BLR cloud models of Table~\ref{tab:agn}. Two calculations are presented
for each model, one including fluorescent excitation (both \lya\ and
\lyb, solid lines) and one not including fluorescent excitation
(dotted lines). At each abundance, the lines in order of increasing flux
correspond to increasing turbulent velocities $0$, $10$, $20$, and $40$
$\rm km\,s^{-1}$.  \label{fig:fe2cog}}
\end{figure}

\begin{figure}[p]
\plotone{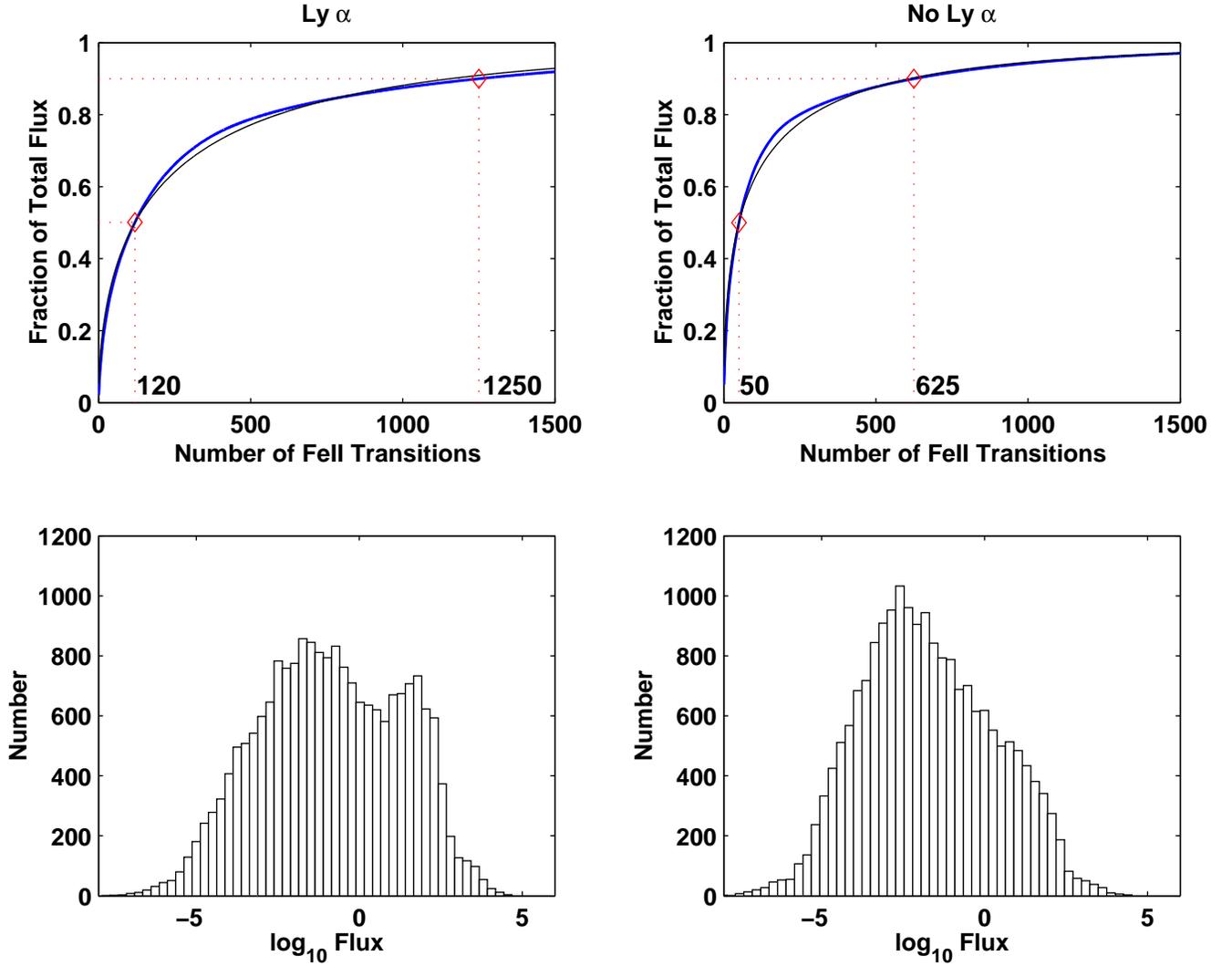}
\figcaption[f5.eps]{The top two panels show the cumulative sum of the
\protect\ion{Fe}{2} flux with (left) and without (right) \lya\ fluorescent
excitation.
The number of transitions required in the
sum to reach 50\% and 90\% of the total flux are as indicated. The bottom
two panels show the distribution of $\log_{10}F$ with (left) and
without (right) \lya\ fluorescent excitation.  BLR model~A was used for all
calculations. \label{fig:flx2cum}}
\end{figure}

\begin{figure}[p]
\epsscale{0.8}
\plotone{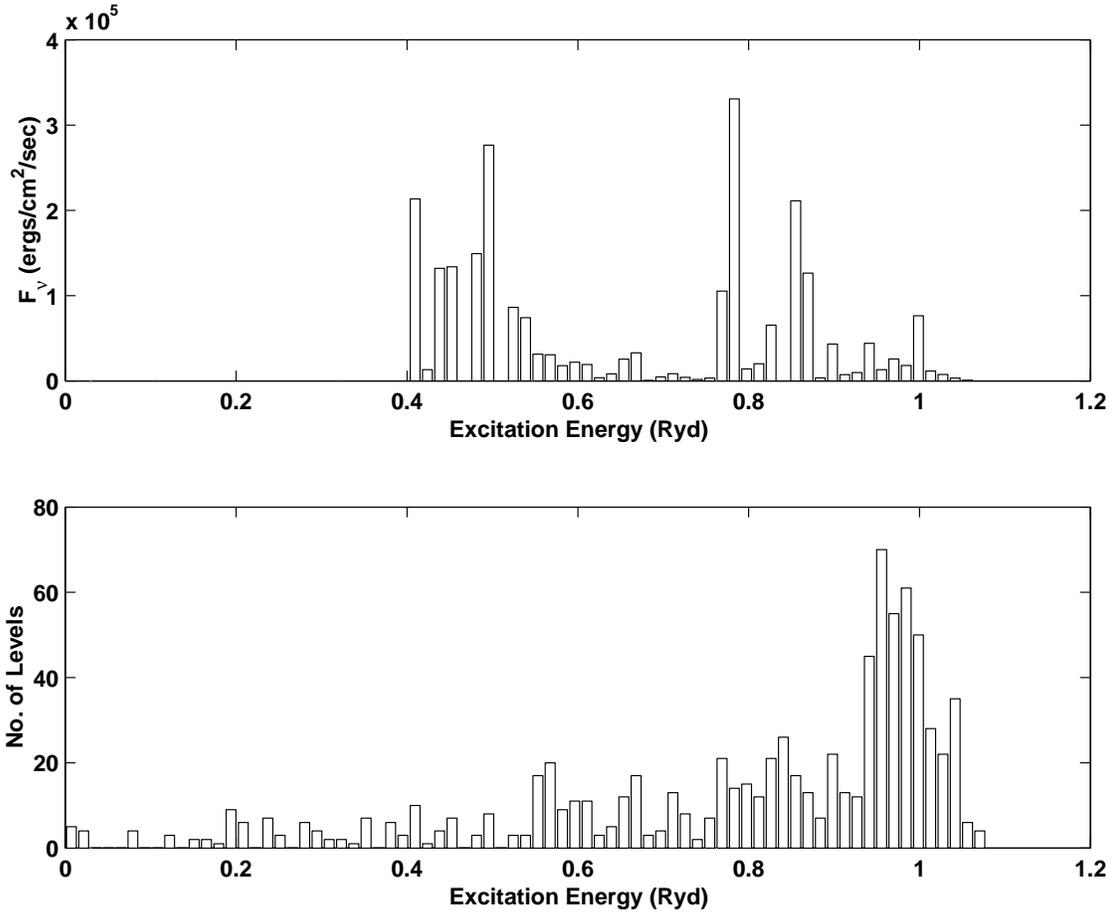}
\figcaption[f6.eps]{The lower panel shows a histogram of the number of included
\protect\ion{Fe}{2} levels within each energy bin. The upper panel
shows the total flux arising from transitions with upper levels in
each bin for BLR model A including fluorescent excitation.
\label{fig:flx2hist}}
\end{figure}

\begin{figure}[p]
\epsscale{1.0}
\plotone{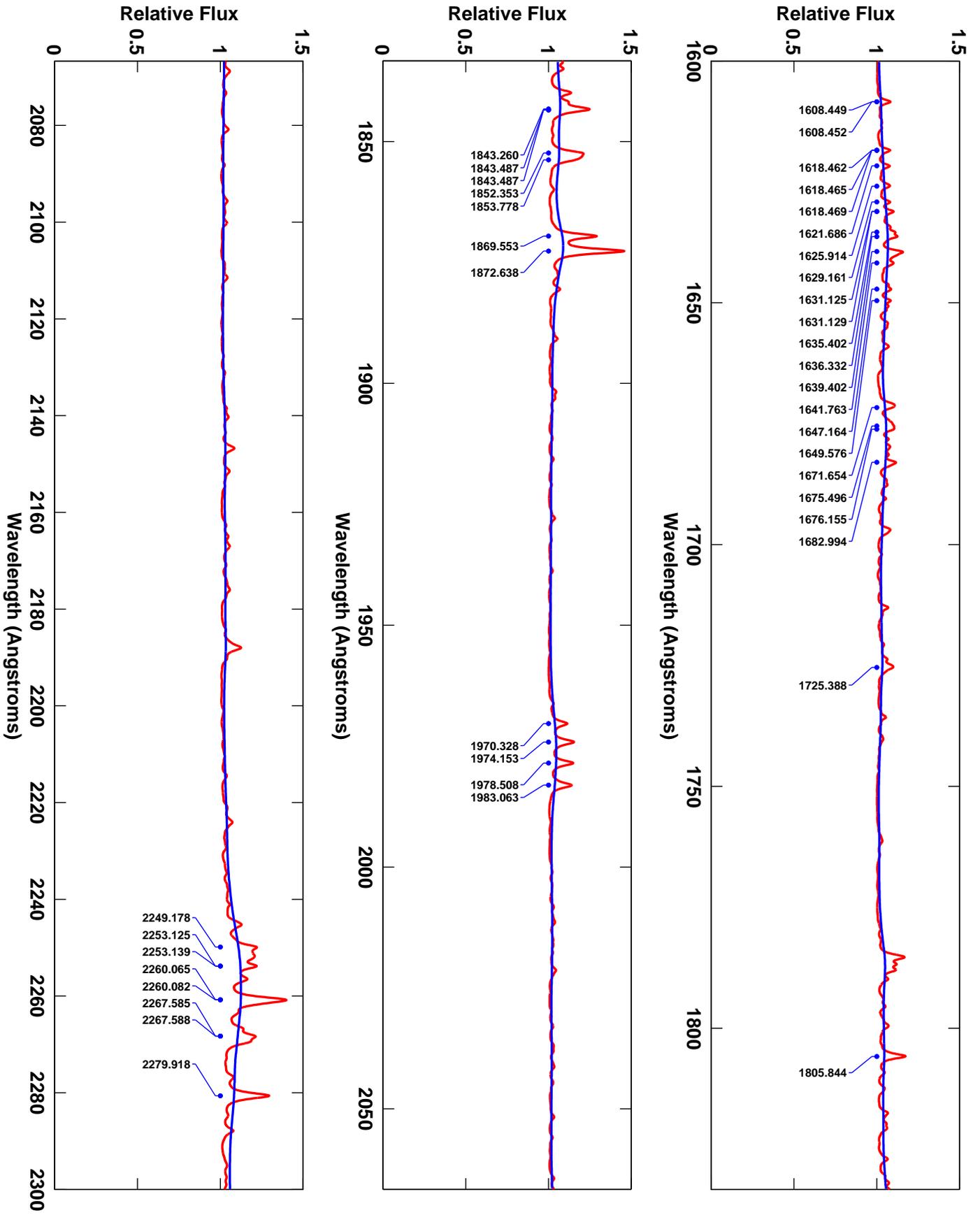}
\figcaption[f7.eps]{The predicted flux and line identifications for
1600-2300~\AA\ for BLR model A with \lya\ and \lyb\ fluorescence.
\label{fig:flx_1600_2300}}
\end{figure}

\begin{figure}[p]
\plotone{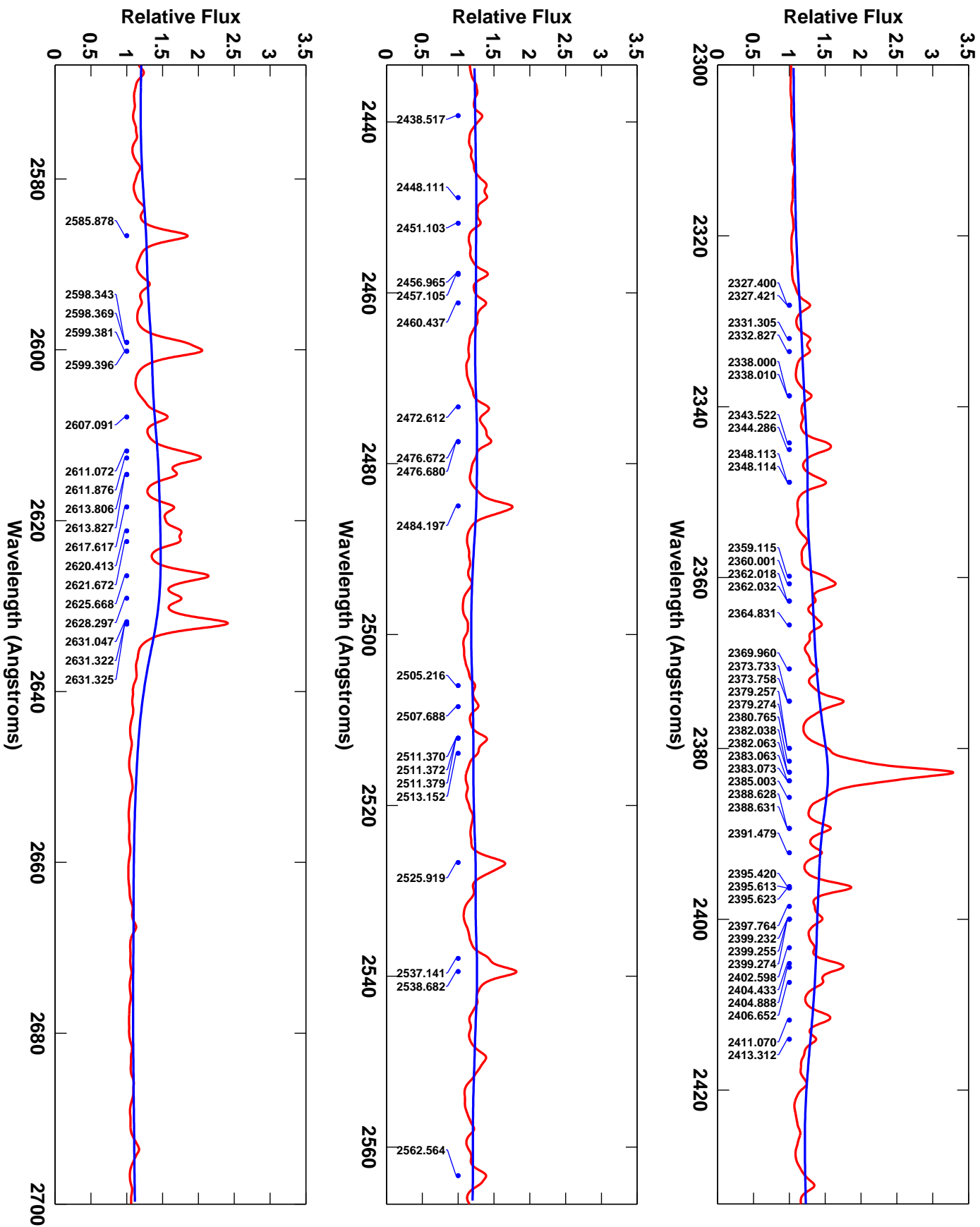}
\figcaption[f8.eps]{The 2300-2700~\AA\ region predicted for AGN model A
with \lya\ and \lyb\ fluorescence.
\label{fig:flx_2300_2700}}
\end{figure}

\begin{figure}[p]
\plotone{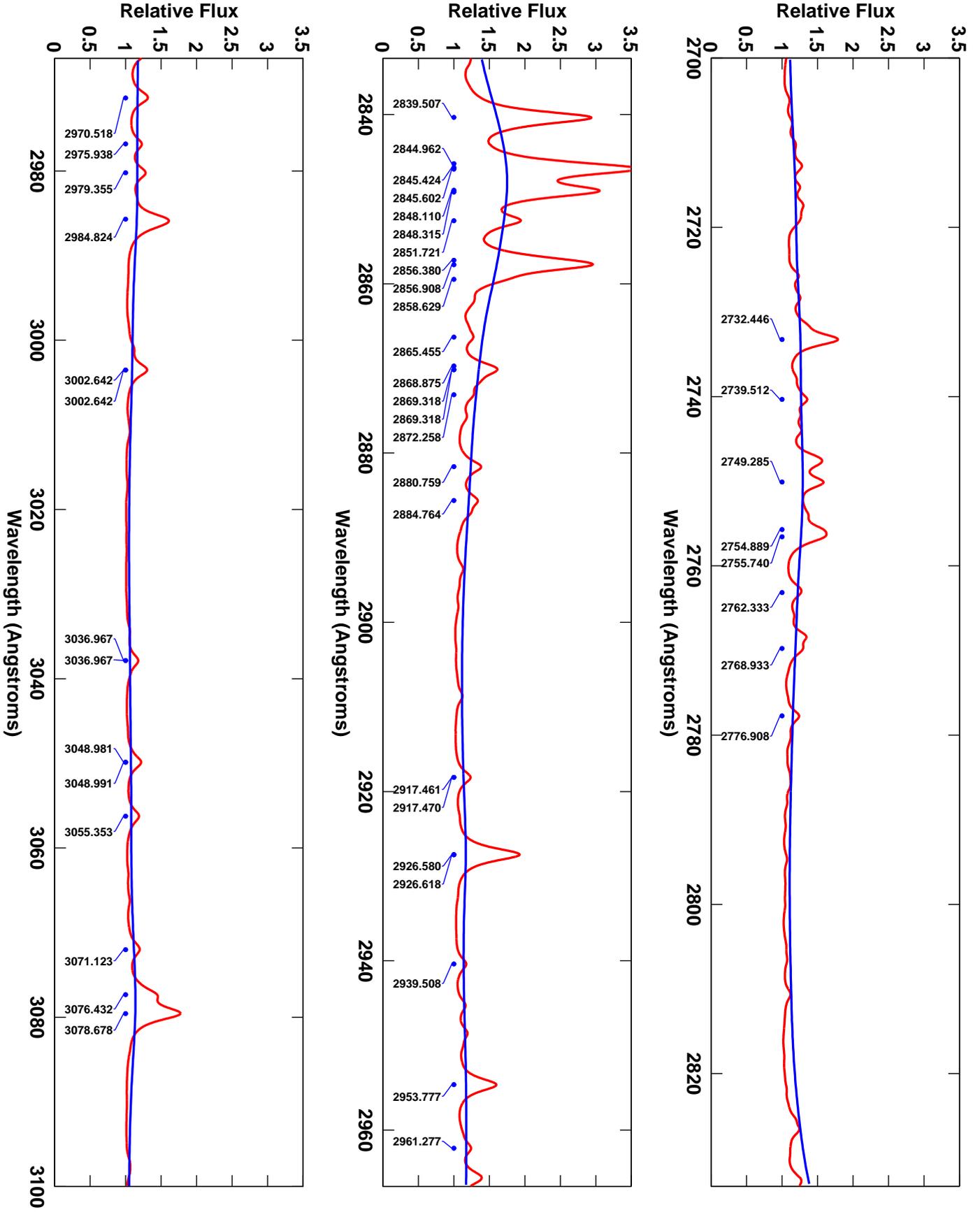}
\figcaption[f9.eps]{The 2700-3100~\AA\ region predicted for AGN model A
with \lya\ and \lyb\ fluorescence.
\label{fig:flx_2700_3100}}
\end{figure}

\begin{figure}[p]
\plotone{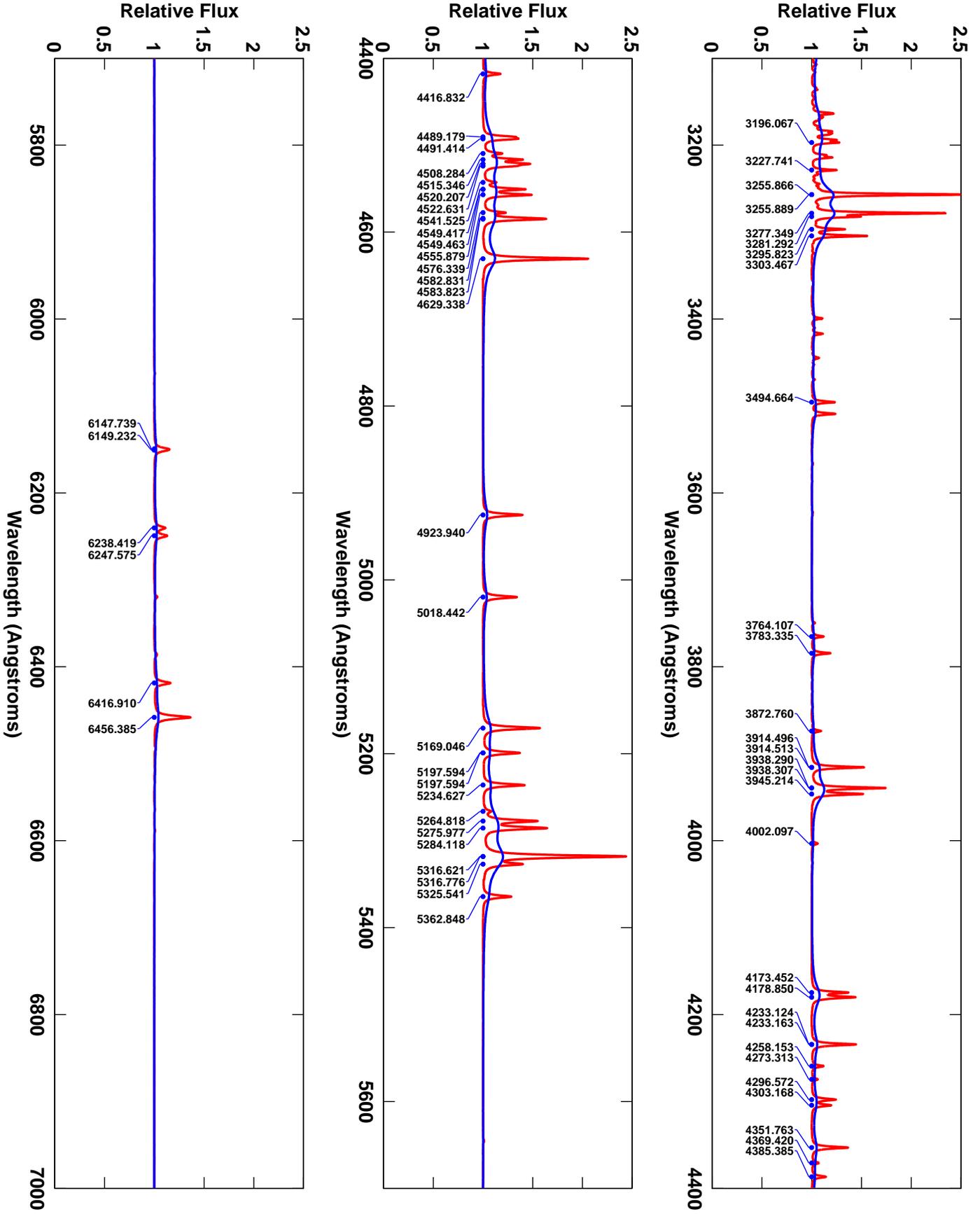}
\figcaption[f10.eps]{The 3100-7000~\AA\ region predicted for AGN model A
with \lya\ and \lyb\ fluorescence.
\label{fig:flx_3100_7000}}
\end{figure}

\begin{figure}[p]
\plotone{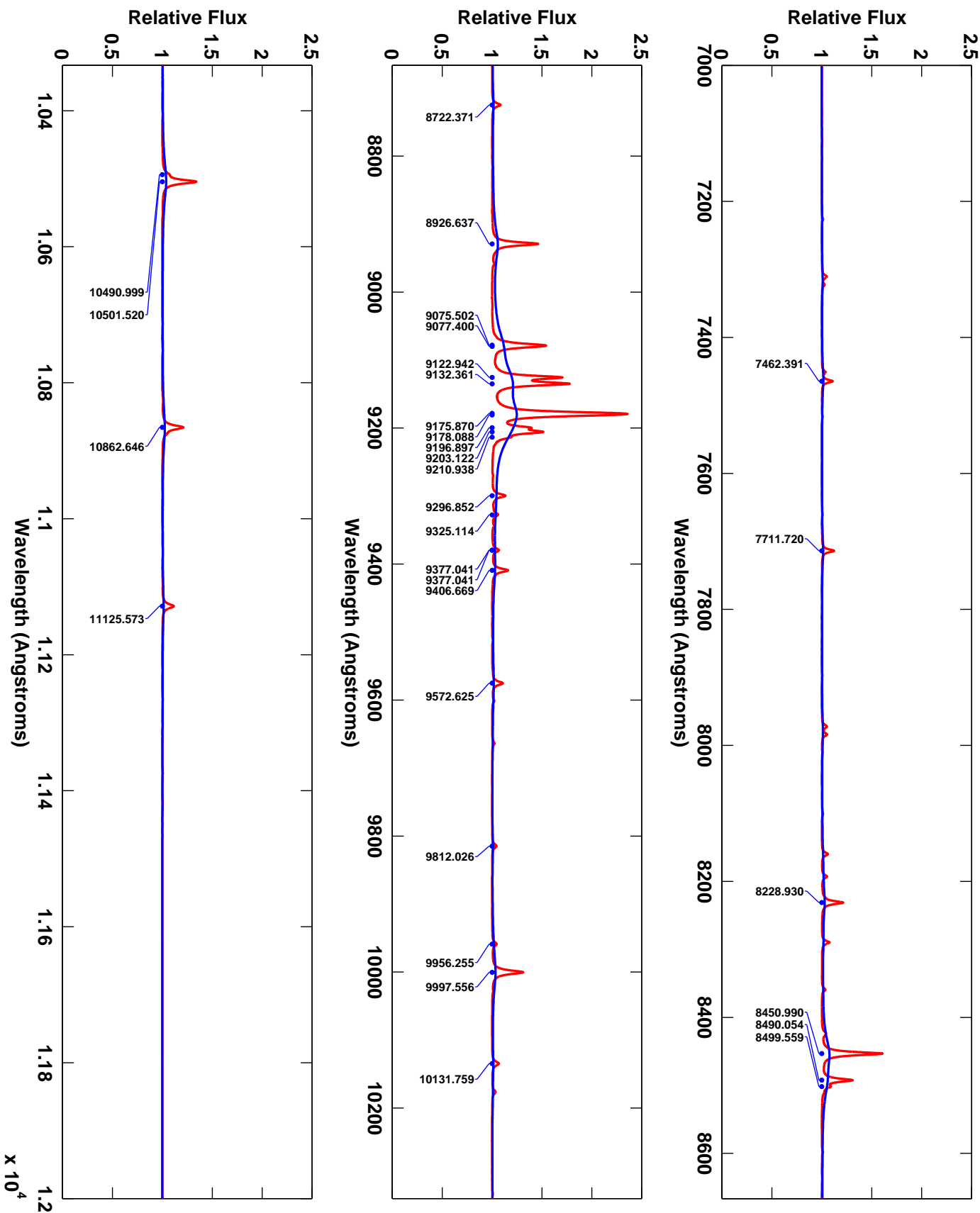}
\figcaption[f11.eps]{The 7000-1200~\AA\ region predicted for AGN model A
with \lya\ and \lyb\ fluorescence.
\label{fig:flx_7000_12000}}
\end{figure}

\begin{figure}[p]
\epsscale{0.8}
\plotone{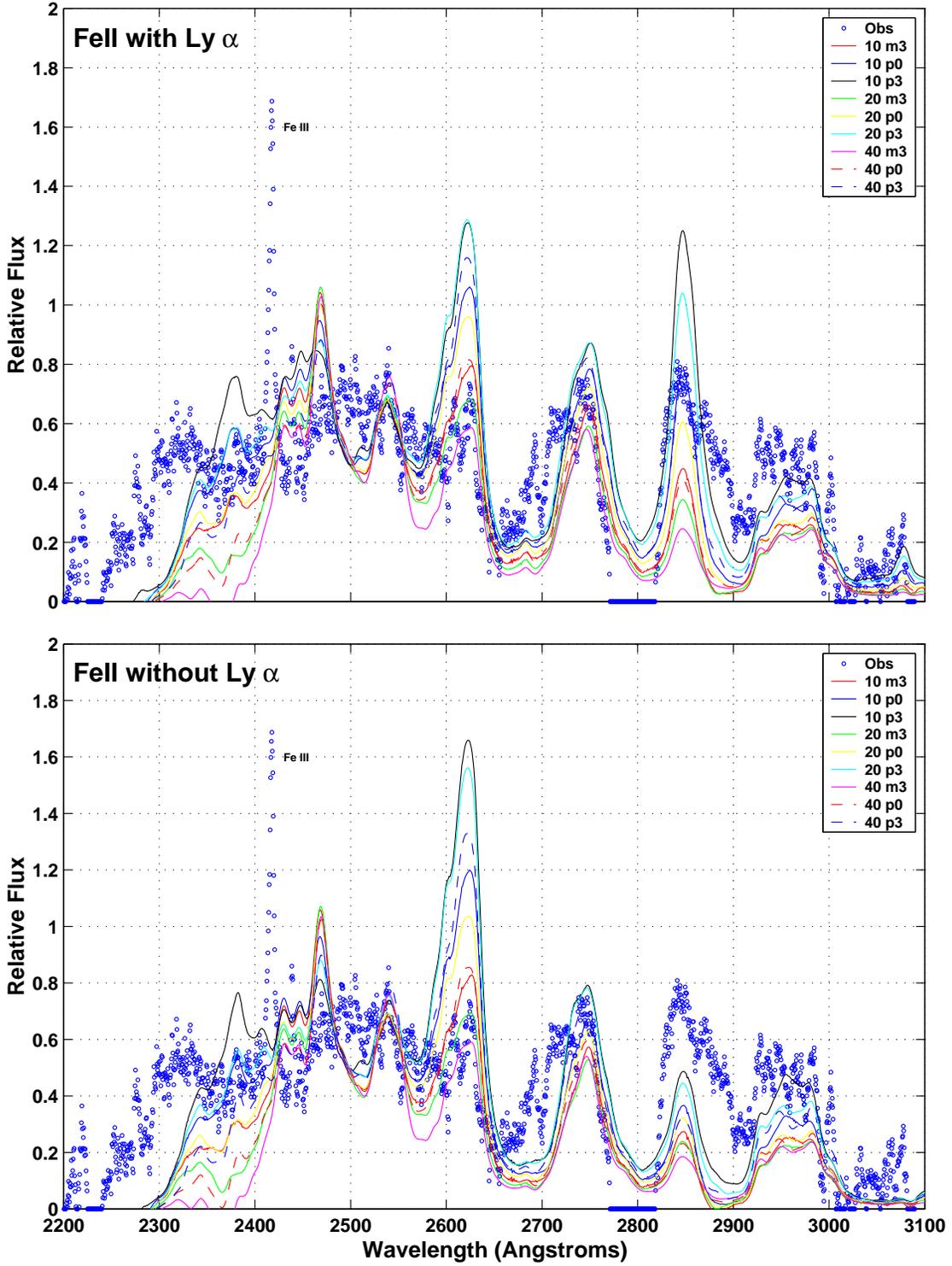}
\figcaption[f12.eps]{A comparison of the predicted UV \ion{Fe}{2} spectrum
corresponding to model~B of Table~\ref{tab:agn} and the empirical UV
\ion{Fe}{2} - \ion{Fe}{3} template of Vestergaard \& Wilkes (2001). The
upper panel includes fluorescent excitation while the lower panel does
not. The various colours and line styles correspond to different
combinations of iron abundance and cloud turbulence velocity as
indicated by the legend in each panel.  \label{fig:1zw1_uv}}
\end{figure}

\begin{figure}[p]
\plotone{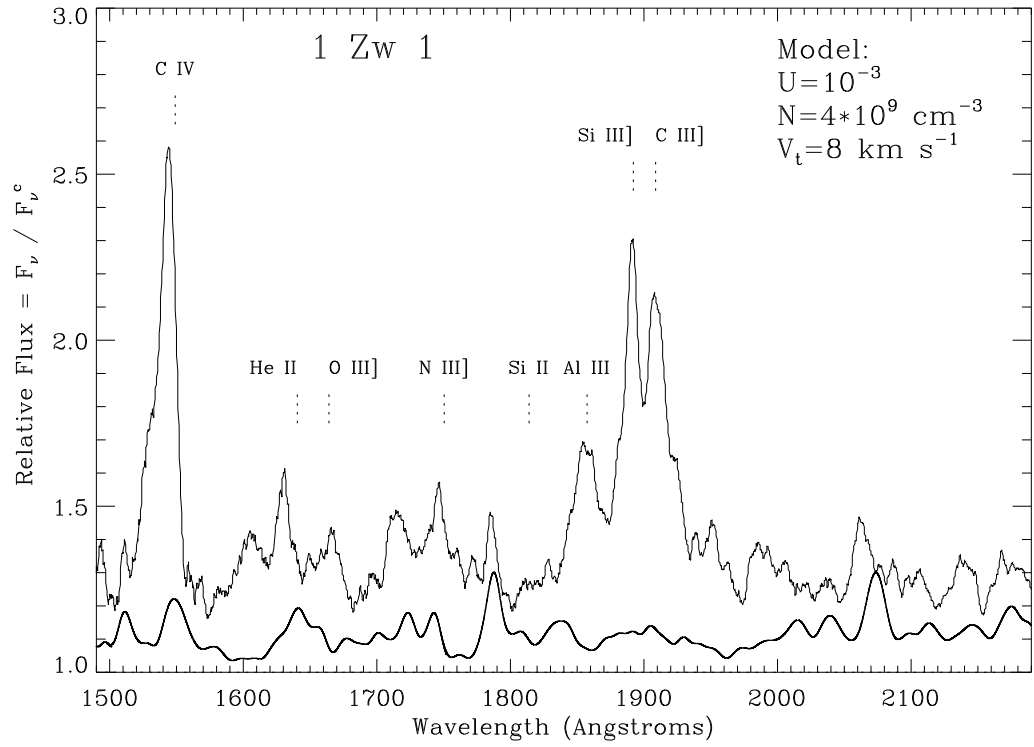}
\figcaption[f13.eps]{The UV spectrum of I~Zw~1 (Marziani \etal\ 1996)
compared to the theoretical \protect\ion{Fe}{2} spectra The observed
spectrum has been smoothed over 15 pixels and rectified to a
power-law continuum ($F^c=11\, (1500\AA/\lambda)^{1.1}$). The
observations are off-set by $+0.1$ for clarity. A few of the prominent
emission lines in this region are identified using Table 4A of Laor
\etal\ (1994). \label{fig:1zw1}}
\end{figure}

\begin{figure}[p]
\plotone{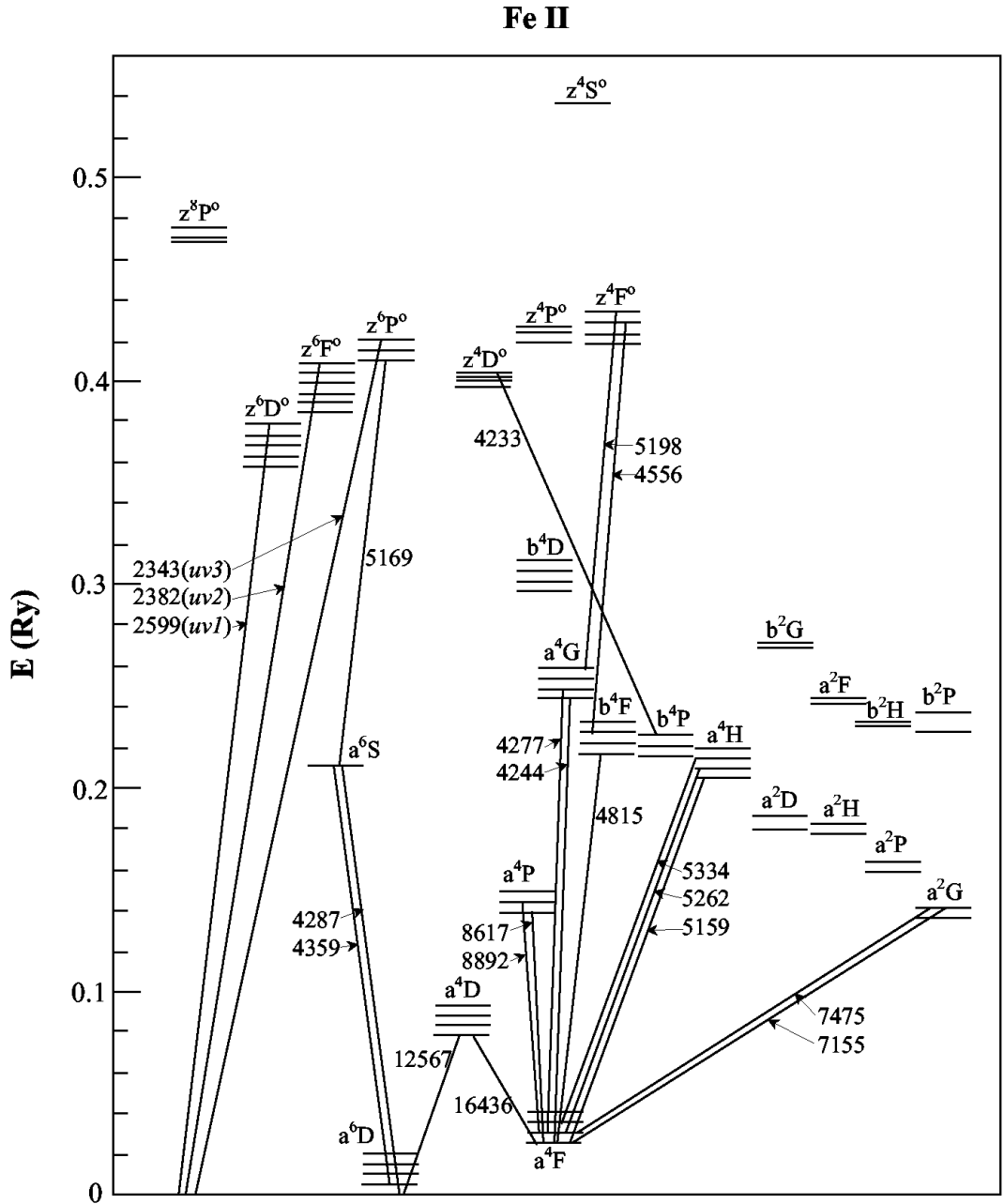}
\figcaption[f14.eps]{A partial Grotrian diagram for \protect\ion{Fe}{2} showing
the low-lying optical and infrared transitions. \label{fig:fe2_lowe}}
\end{figure}

\begin{figure}[p]
\plotone{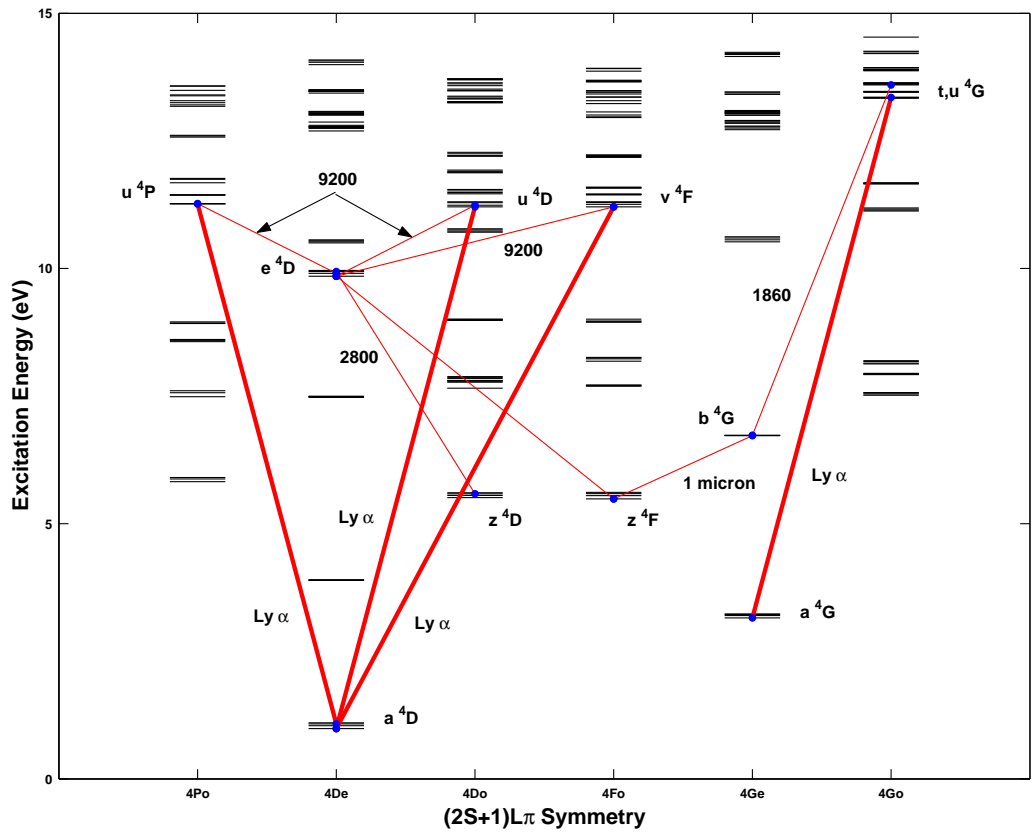}
\figcaption[f15.eps]{The $1\,$\mum\ IR lines and the $\lambda\,9220\,$\AA\
features in the \protect\ion{Fe}{2} quartet system, showing the
\protect\lya\ fluorescence and cascade routes.  The $1\,$\mum\ lines
correspond to multiplet transitions b$^4G$ - z($^4F^o,^4D^o$) given at
the end of Table 11; see the text for details.
\label{fig:atom_fe2_lyair}}
\end{figure}

\clearpage
\newpage

\begin{deluxetable}{rrcrrr}
\tablewidth{0pt}
\tablecaption{\protect\ion{Fe}{2} transitions from $\rm a\,^4\!D^e$
within $\pm3$\AA\ of \lya. \label{tab:lyalines}}
\tablecolumns{6}
\tablehead{
\colhead{$\lambda$} & \colhead{$\Delta\lambda/\Delta\lambda_D$} & \colhead{Transition} &
\multicolumn{3}{c}{\hrulefill\,$\log(gf)$\hrulefill} \\
\colhead{(\AA)} & \colhead{~} & \colhead{$l\rightarrow\,u$} & \colhead{N96} & \colhead{K91} & \colhead{K81}
}
\startdata
1212.97 & $-51.9$ & $\rm a^4D^e_{7/2} - u^4D^o_{7/2}$ & $-1.78$ & $-1.47$ & $-2.39$ \\
1213.13 & $-48.7$ & $\rm a^4D^e_{7/2} - v^4F^o_{9/2}$ & $-0.72$ & $-1.18$ & $-1.70$ \\
1213.74 & $-37.1$ & $\rm a^4D^e_{3/2} - u^4F^o_{5/2}$ & $-1.07$ & $-1.31$ & $-2.16$ \\
1213.76 & $-36.7$ & $\rm a^4D^e_{5/2} - u^4F^o_{7/2}$ & $-0.89$ & $-1.17$ & $-1.95$ \\
1214.07 & $-30.8$ & $\rm a^4D^e_{3/2} - u^4D^o_{3/2}$ & $-2.42$ & $-1.73$ & $-3.67$ \\
1214.15 & $-29.2$ & $\rm a^4D^e_{1/2} - v^4F^o_{3/2}$ & $-1.28$ & $-1.49$ & $-2.39$ \\
1214.29 & $-26.6$ & $\rm a^4D^e_{1/2} - u^4D^o_{1/2}$ & $-2.63$ & $-2.16$ & $-3.25$ \\
1214.40 & $-24.4$ & $\rm a^4D^e_{7/2} - x^6F^o_{7/2}$ & $\null$ & $-2.23$ & $-2.40$ \\
\enddata
\tablecomments{$\Delta\lambda_D$ is the Doppler width of \protect\lya\ at
$10^4\,$K. The references for the $\log(gf)$ are N96 (Nahar 1996),
K91 (Kurucz 1991), and K81 (Kurucz 1981).
The shift and $\sigma$ entries refer to the average differences
and standard deviations relative to N96.
This list is only the closest lines in the models including
\protect\lya\ pumping.}
\end{deluxetable}

\begin{deluxetable}{ccrr}
\tablewidth{0pt}
\tablecaption{The four AGN cloud models considered in this work.
\label{tab:agn}}
\tablecolumns{4}
\tablehead{
\colhead{Model} & \colhead{$U_{\rm ion}$} & \colhead{$\log_{10} N_H$} & \colhead{$\log_{10} F_{H\beta}$}
}
\startdata
A  & $10^{-2}$ & $9.6$ & 5.68 \\
B  & $10^{-2}$ & $11.6$ & 7.20 \\
C  & $10^{-3}$ & $9.6$ & 4.55 \\
D  & $10^{-3}$ & $11.6$ & 6.70 \\
\enddata
\tablecomments{The shape of the photoionizing continuum was taken from
Mathews \& Ferland (1988).}
\end{deluxetable}

%
% problems in the format of this table!
\begin{deluxetable}{rllrr||rllrr}
\footnotesize
\tablewidth{18cm}
\tablecaption{\protect\ion{Fe}{2} fluxes for the model~A with \protect\lya\ and \protect\lyb\ pumping.}
\tablehead{
\colhead{$\lambda$(\AA)} & \colhead{Upper} & \colhead{Lower} & \colhead{$E_{l}$(eV)} & \colhead{Flux} &
\colhead{$\lambda$(\AA)} & \colhead{Upper} & \colhead{Lower} & \colhead{$E_{l}$(eV)} & \colhead{Flux}
}
\startdata
927.174 & $u\,^{6}D^{o}_{7/2}$ & $a\,^{6}D^{e}_{7/2}$ & 0.048 & $9.89\,^{+ 2}$ & 1309.555 & $x\,^{2}I^{o}_{11/2}$ & $a\,^{2}I^{e}_{11/2}$ & 4.079 & $7.38\,^{+ 2}$ \\
1133.674 & $u\,^{4}D^{o}_{5/2}$ & $a\,^{4}F^{e}_{7/2}$ & 0.301 & $5.59\,^{+ 3}$ & 1310.589 & $t\,^{2}H^{o}_{11/2}$ & $b\,^{2}G^{e}_{9/2}$ & 3.766 & $7.31\,^{+ 2}$ \\
1135.578 & $u\,^{4}P^{o}_{3/2}$ & $a\,^{4}F^{e}_{5/2}$ & 0.352 & $5.18\,^{+ 2}$ & 1311.062 & $x\,^{2}I^{o}_{13/2}$ & $a\,^{2}I^{e}_{13/2}$ & 4.074 & $7.30\,^{+ 2}$ \\
1138.940 & $u\,^{4}D^{o}_{5/2}$ & $a\,^{4}F^{e}_{5/2}$ & 0.352 & $2.43\,^{+ 3}$ & 1311.615 & $t\,^{2}H^{o}_{9/2}$ & $b\,^{2}G^{e}_{7/2}$ & 3.813 & $5.48\,^{+ 2}$ \\
1144.049 & $u\,^{6}D^{o}_{7/2}$ & $b\,^{4}P^{e}_{5/2}$ & 2.582 & $2.28\,^{+ 3}$ & 1312.490 & $u\,^{4}G^{o}_{9/2}$ & $b\,^{4}D^{e}_{7/2}$ & 3.902 & $1.33\,^{+ 3}$ \\
1159.503 & $u\,^{4}G^{o}_{9/2}$ & $a\,^{4}H^{e}_{11/2}$ & 2.656 & $6.04\,^{+ 2}$ & 1332.731 & $v\,^{6}F^{o}_{3/2}$ & $b\,^{4}D^{e}_{3/2}$ & 3.887 & $6.42\,^{+ 2}$ \\
1170.562 & $u\,^{6}D^{o}_{7/2}$ & $b\,^{4}F^{e}_{7/2}$ & 2.827 & $1.15\,^{+ 3}$ & 1352.735 & $r\,^{4}F^{o}_{9/2}$ & $b\,^{4}D^{e}_{7/2}$ & 3.902 & $1.20\,^{+ 3}$ \\
1175.957 & $u\,^{4}G^{o}_{9/2}$ & $b\,^{4}F^{e}_{9/2}$ & 2.805 & $2.89\,^{+ 3}$ & 1354.748 & $t\,^{2}H^{o}_{11/2}$ & $a\,^{2}I^{e}_{13/2}$ & 4.074 & $6.39\,^{+ 2}$ \\
\enddata
\end{deluxetable}

\clearpage
\addtocounter{table}{-1}

\begin{deluxetable}{rllrr||rllrr}
\footnotesize
\tablewidth{18cm}
\tablecaption{\protect\ion{Fe}{2} fluxes for the model~A with \protect\lya\ and \protect\lyb\ pumping.}
\tablehead{
\colhead{$\lambda$(\AA)} & \colhead{Upper} & \colhead{Lower} & \colhead{$E_{l}$(eV)} & \colhead{Flux} &
\colhead{$\lambda$(\AA)} & \colhead{Upper} & \colhead{Lower} & \colhead{$E_{l}$(eV)} & \colhead{Flux}
}
\startdata
1538.632 & $u\,^{4}D^{o}_{3/2}$ & $a\,^{4}G^{e}_{5/2}$ & 3.229 & $1.77\,^{+ 3}$ & 1723.684 & $n\,^{4}F^{o}_{3/2}$ & $b\,^{4}G^{e}_{5/2}$ & 6.727 & $1.20\,^{+ 3}$ \\
1539.045 & $v\,^{4}F^{o}_{7/2}$ & $a\,^{4}G^{e}_{9/2}$ & 3.198 & $4.11\,^{+ 3}$ & 1725.388 & $n\,^{4}F^{o}_{7/2}$ & $b\,^{4}G^{e}_{9/2}$ & 6.726 & $1.60\,^{+ 3}$ \\
1539.461 & $v\,^{4}F^{o}_{9/2}$ & $a\,^{4}G^{e}_{11/2}$ & 3.151 & $3.12\,^{+ 3}$ & 1735.681 & $n\,^{4}F^{o}_{9/2}$ & $b\,^{4}G^{e}_{11/2}$ & 6.721 & $1.23\,^{+ 3}$ \\
1540.369 & $v\,^{4}P^{o}_{5/2}$ & $a\,^{4}G^{e}_{7/2}$ & 3.220 & $1.28\,^{+ 3}$ & 1739.541 & $u\,^{4}G^{o}_{9/2}$ & $c\,^{4}F^{e}_{7/2}$ & 6.220 & $5.04\,^{+ 2}$ \\
1543.234 & $v\,^{4}F^{o}_{7/2}$ & $a\,^{4}G^{e}_{7/2}$ & 3.220 & $1.17\,^{+ 3}$ & 1779.499 & $m\,^{4}D^{o}_{7/2}$ & $b\,^{4}G^{e}_{9/2}$ & 6.726 & $5.03\,^{+ 2}$ \\
1547.064 & $x\,^{6}F^{o}_{9/2}$ & $a\,^{4}G^{e}_{11/2}$ & 3.151 & $1.42\,^{+ 3}$ & 1784.891 & $n\,^{4}F^{o}_{5/2}$ & $b\,^{4}G^{e}_{7/2}$ & 6.727 & $1.52\,^{+ 3}$ \\
1550.499 & $x\,^{6}F^{o}_{7/2}$ & $a\,^{4}G^{e}_{9/2}$ & 3.198 & $5.37\,^{+ 2}$ & 1785.272 & $x\,^{6}P^{o}_{7/2}$ & $a\,^{6}S^{e}_{5/2}$ & 2.890 & $2.00\,^{+ 3}$ \\
1574.497 & $t\,^{4}P^{o}_{1/2}$ & $b\,^{4}D^{e}_{1/2}$ & 3.888 & $5.30\,^{+ 2}$ & 1785.485 & $o\,^{4}F^{o}_{5/2}$ & $b\,^{4}G^{e}_{7/2}$ & 6.727 & $1.25\,^{+ 3}$ \\
\enddata
\end{deluxetable}

\clearpage
\addtocounter{table}{-1}

\begin{deluxetable}{rllrr||rllrr}
\footnotesize
\tablewidth{18cm}
\tablecaption{\protect\ion{Fe}{2} fluxes for the model~A with \protect\lya and \protect\lyb pumping.}
\tablehead{
\colhead{$\lambda$(\AA)} & \colhead{Upper} & \colhead{Lower} & \colhead{$E_{l}$(eV)} & \colhead{Flux} &
\colhead{$\lambda$(\AA)} & \colhead{Upper} & \colhead{Lower} & \colhead{$E_{l}$(eV)} & \colhead{Flux}
}
\startdata
1852.353 & $p\,^{4}F^{o}_{9/2}$ & $b\,^{4}G^{e}_{11/2}$ & 6.721 & $1.14\,^{+ 3}$ & 2260.083 & $z\,^{4}F^{o}_{9/2}$ & $a\,^{6}D^{e}_{9/2}$ & 0.000 & $6.95\,^{+ 3}$ \\
1853.028 & $u\,^{6}D^{o}_{7/2}$ & $b\,^{4}G^{e}_{9/2}$ & 6.726 & $1.91\,^{+ 3}$ & 2260.237 & $z\,^{4}D^{o}_{1/2}$ & $a\,^{6}D^{e}_{1/2}$ & 0.121 & $6.09\,^{+ 2}$ \\
1853.355 & $u\,^{6}D^{o}_{7/2}$ & $b\,^{4}G^{e}_{7/2}$ & 6.727 & $1.22\,^{+ 3}$ & 2260.861 & $z\,^{4}F^{o}_{5/2}$ & $a\,^{6}D^{e}_{3/2}$ & 0.107 & $1.56\,^{+ 3}$ \\
1853.778 & $p\,^{4}F^{o}_{9/2}$ & $b\,^{4}G^{e}_{9/2}$ & 6.726 & $6.54\,^{+ 2}$ & 2262.686 & $z\,^{4}D^{o}_{3/2}$ & $a\,^{6}D^{e}_{3/2}$ & 0.107 & $8.07\,^{+ 2}$ \\
1869.553 & $u\,^{4}G^{o}_{11/2}$ & $b\,^{4}G^{e}_{11/2}$ & 6.721 & $8.36\,^{+ 3}$ & 2265.994 & $z\,^{4}D^{o}_{5/2}$ & $a\,^{6}D^{e}_{5/2}$ & 0.083 & $1.20\,^{+ 3}$ \\
1872.638 & $u\,^{4}G^{o}_{9/2}$ & $b\,^{4}G^{e}_{9/2}$ & 6.726 & $1.27\,^{+ 4}$ & 2267.584 & $z\,^{4}F^{o}_{7/2}$ & $a\,^{6}D^{e}_{5/2}$ & 0.083 & $3.17\,^{+ 3}$ \\
1872.973 & $u\,^{4}G^{o}_{9/2}$ & $b\,^{4}G^{e}_{7/2}$ & 6.727 & $9.86\,^{+ 2}$ & 2268.562 & $z\,^{4}D^{o}_{3/2}$ & $a\,^{6}D^{e}_{1/2}$ & 0.121 & $5.93\,^{+ 2}$ \\
1877.716 & $u\,^{4}G^{o}_{7/2}$ & $b\,^{4}G^{e}_{9/2}$ & 6.726 & $5.18\,^{+ 2}$ & 2268.822 & $z\,^{4}D^{o}_{7/2}$ & $a\,^{6}D^{e}_{7/2}$ & 0.048 & $2.31\,^{+ 3}$ \\
\enddata
\end{deluxetable}

\clearpage
\addtocounter{table}{-1}

\begin{deluxetable}{rllrr||rllrr}
\footnotesize
\tablewidth{18cm}
\tablecaption{\protect\ion{Fe}{2} fluxes for the model~A with \protect\lya\ and \protect\lyb\ pumping.}
\tablehead{
\colhead{$\lambda$(\AA)} & \colhead{Upper} & \colhead{Lower} & \colhead{$E_{l}$(eV)} & \colhead{Flux} &
\colhead{$\lambda$(\AA)} & \colhead{Upper} & \colhead{Lower} & \colhead{$E_{l}$(eV)} & \colhead{Flux}
}
\startdata
2356.941 & $u\,^{2}G^{o}_{7/2}$ & $c\,^{4}F^{e}_{7/2}$ & 6.220 & $5.50\,^{+ 2}$ & 2384.383 & $h\,^{4}G^{e}_{5/2}$ & $x\,^{4}F^{o}_{3/2}$ & 8.255 & $5.14\,^{+ 2}$ \\
2359.106 & $z\,^{6}P^{o}_{5/2}$ & $a\,^{6}D^{e}_{3/2}$ & 0.107 & $2.72\,^{+ 3}$ & 2384.385 & $z\,^{4}D^{o}_{3/2}$ & $a\,^{4}F^{e}_{3/2}$ & 0.386 & $1.70\,^{+ 3}$ \\
2359.108 & $a\,^{6}G^{e}_{7/2}$ & $z\,^{6}F^{o}_{5/2}$ & 5.247 & $9.48\,^{+ 2}$ & 2385.003 & $z\,^{4}F^{o}_{7/2}$ & $a\,^{4}F^{e}_{5/2}$ & 0.352 & $3.54\,^{+ 3}$ \\
2360.001 & $z\,^{4}F^{o}_{9/2}$ & $a\,^{4}F^{e}_{9/2}$ & 0.232 & $8.16\,^{+ 3}$ & 2386.376 & $e\,^{4}F^{e}_{9/2}$ & $z\,^{4}F^{o}_{9/2}$ & 5.482 & $9.60\,^{+ 2}$ \\
2360.292 & $z\,^{4}D^{o}_{5/2}$ & $a\,^{4}F^{e}_{7/2}$ & 0.301 & $2.37\,^{+ 3}$ & 2388.631 & $z\,^{6}F^{o}_{7/2}$ & $a\,^{6}D^{e}_{7/2}$ & 0.048 & $5.26\,^{+ 3}$ \\
2361.738 & $a\,^{6}G^{e}_{11/2}$ & $z\,^{6}F^{o}_{11/2}$ & 5.201 & $5.96\,^{+ 2}$ & 2389.311 & $h\,^{4}G^{e}_{9/2}$ & $x\,^{4}F^{o}_{7/2}$ & 8.226 & $1.70\,^{+ 3}$ \\
2362.018 & $z\,^{4}F^{o}_{7/2}$ & $a\,^{4}F^{e}_{7/2}$ & 0.301 & $4.41\,^{+ 3}$ & 2391.480 & $z\,^{4}F^{o}_{9/2}$ & $a\,^{4}F^{e}_{7/2}$ & 0.301 & $7.02\,^{+ 3}$ \\
2363.859 & $a\,^{6}G^{e}_{9/2}$ & $z\,^{6}F^{o}_{7/2}$ & 5.235 & $2.74\,^{+ 3}$ & 2391.739 & $u\,^{4}P^{o}_{3/2}$ & $c\,^{4}P^{e}_{1/2}$ & 6.085 & $7.70\,^{+ 2}$ \\
\enddata
\end{deluxetable}

\clearpage
\addtocounter{table}{-1}

\begin{deluxetable}{rllrr||rllrr}
\footnotesize
\tablewidth{18cm}
\tablecaption{\protect\ion{Fe}{2} fluxes for the model~A with \protect\lya\ and \protect\lyb\ pumping.}
\tablehead{
\colhead{$\lambda$(\AA)} & \colhead{Upper} & \colhead{Lower} & \colhead{$E_{l}$(eV)} & \colhead{Flux} &
\colhead{$\lambda$(\AA)} & \colhead{Upper} & \colhead{Lower} & \colhead{$E_{l}$(eV)} & \colhead{Flux}
}
\startdata
2430.462 & $u\,^{4}D^{o}_{5/2}$ & $c\,^{4}P^{e}_{3/2}$ & 6.136 & $2.85\,^{+ 3}$ & 2476.680 & $z\,^{6}F^{o}_{7/2}$ & $a\,^{4}F^{e}_{9/2}$ & 0.232 & $4.18\,^{+ 3}$ \\
2430.994 & $y\,^{4}S^{o}_{3/2}$ & $c\,^{4}P^{e}_{3/2}$ & 6.136 & $1.02\,^{+ 3}$ & 2482.118 & $z\,^{4}I^{o}_{13/2}$ & $a\,^{4}H^{e}_{13/2}$ & 2.634 & $8.35\,^{+ 2}$ \\
2438.517 & $v\,^{4}F^{o}_{3/2}$ & $c\,^{4}P^{e}_{5/2}$ & 6.223 & $5.14\,^{+ 3}$ & 2482.659 & $x\,^{4}G^{o}_{9/2}$ & $a\,^{4}G^{e}_{11/2}$ & 3.151 & $6.57\,^{+ 2}$ \\
2438.602 & $v\,^{4}F^{o}_{5/2}$ & $c\,^{4}F^{e}_{3/2}$ & 6.206 & $6.01\,^{+ 2}$ & 2482.666 & $r\,^{2}F^{o}_{7/2}$ & $d\,^{2}G^{e}_{9/2}$ & 7.266 & $5.34\,^{+ 2}$ \\
2439.300 & $y\,^{4}H^{o}_{13/2}$ & $a\,^{4}G^{e}_{11/2}$ & 3.151 & $7.22\,^{+ 2}$ & 2482.869 & $e\,^{4}D^{e}_{3/2}$ & $z\,^{4}D^{o}_{5/2}$ & 5.550 & $6.22\,^{+ 2}$ \\
2439.444 & $d\,^{4}D^{e}_{7/2}$ & $z\,^{6}D^{o}_{9/2}$ & 4.766 & $7.05\,^{+ 2}$ & 2484.197 & $z\,^{6}F^{o}_{9/2}$ & $a\,^{4}F^{e}_{9/2}$ & 0.232 & $1.21\,^{+ 4}$ \\
2439.928 & $u\,^{4}D^{o}_{3/2}$ & $c\,^{4}F^{e}_{3/2}$ & 6.206 & $9.69\,^{+ 2}$ & 2484.447 & $e\,^{4}D^{e}_{7/2}$ & $z\,^{4}D^{o}_{7/2}$ & 5.508 & $1.95\,^{+ 3}$ \\
2441.131 & $c\,^{4}G^{e}_{9/2}$ & $z\,^{4}F^{o}_{9/2}$ & 5.482 & $1.02\,^{+ 3}$ & 2485.010 & $v\,^{4}F^{o}_{9/2}$ & $c\,^{4}F^{e}_{9/2}$ & 6.216 & $1.75\,^{+ 3}$ \\
\enddata
\end{deluxetable}

\clearpage
\addtocounter{table}{-1}

\begin{deluxetable}{rllrr||rllrr}
\footnotesize
\tablewidth{18cm}
\tablecaption{\protect\ion{Fe}{2} fluxes for the model~A with \protect\lya\ and \protect\lyb\ pumping.}
\tablehead{
\colhead{$\lambda$(\AA)} & \colhead{Upper} & \colhead{Lower} & \colhead{$E_{l}$(eV)} & \colhead{Flux} &
\colhead{$\lambda$(\AA)} & \colhead{Upper} & \colhead{Lower} & \colhead{$E_{l}$(eV)} & \colhead{Flux}
}
\startdata
2536.844 & $z\,^{4}H^{o}_{13/2}$ & $a\,^{4}H^{e}_{11/2}$ & 2.656 & $8.60\,^{+ 2}$ & 2669.930 & $v\,^{4}D^{o}_{3/2}$ & $c\,^{4}P^{e}_{1/2}$ & 6.085 & $5.70\,^{+ 2}$ \\
2537.141 & $b\,^{6}D^{e}_{9/2}$ & $z\,^{6}D^{o}_{9/2}$ & 4.766 & $3.43\,^{+ 3}$ & 2681.041 & $v\,^{4}D^{o}_{1/2}$ & $c\,^{4}P^{e}_{1/2}$ & 6.085 & $9.21\,^{+ 2}$ \\
2538.682 & $b\,^{6}D^{e}_{7/2}$ & $z\,^{6}D^{o}_{5/2}$ & 4.816 & $1.09\,^{+ 4}$ & 2692.604 & $y\,^{2}H^{o}_{11/2}$ & $b\,^{2}G^{e}_{9/2}$ & 3.766 & $5.10\,^{+ 2}$ \\
2538.993 & $z\,^{4}G^{o}_{11/2}$ & $a\,^{4}H^{e}_{13/2}$ & 2.634 & $8.09\,^{+ 2}$ & 2692.834 & $z\,^{4}F^{o}_{5/2}$ & $a\,^{4}D^{e}_{7/2}$ & 0.986 & $2.56\,^{+ 3}$ \\
2542.320 & $z\,^{6}F^{o}_{1/2}$ & $a\,^{4}F^{e}_{3/2}$ & 0.386 & $7.62\,^{+ 2}$ & 2699.194 & $v\,^{4}D^{o}_{3/2}$ & $c\,^{4}P^{e}_{3/2}$ & 6.136 & $6.40\,^{+ 2}$ \\
2543.374 & $z\,^{4}H^{o}_{11/2}$ & $a\,^{4}H^{e}_{9/2}$ & 2.675 & $6.08\,^{+ 2}$ & 2709.381 & $z\,^{4}F^{o}_{3/2}$ & $a\,^{4}D^{e}_{5/2}$ & 1.040 & $1.83\,^{+ 3}$ \\
2550.153 & $b\,^{6}D^{e}_{9/2}$ & $z\,^{6}D^{o}_{7/2}$ & 4.791 & $1.20\,^{+ 3}$ & 2710.550 & $v\,^{4}D^{o}_{1/2}$ & $c\,^{4}P^{e}_{3/2}$ & 6.136 & $9.54\,^{+ 2}$ \\
2562.535 & $z\,^{4}P^{o}_{5/2}$ & $a\,^{4}D^{e}_{7/2}$ & 0.986 & $3.64\,^{+ 3}$ & 2711.841 & $z\,^{2}I^{o}_{13/2}$ & $a\,^{4}G^{e}_{11/2}$ & 3.151 & $1.07\,^{+ 3}$ \\
\enddata
\end{deluxetable}

\clearpage
\addtocounter{table}{-1}

\begin{deluxetable}{rllrr||rllrr}
\footnotesize
\tablewidth{18cm}
\tablecaption{\protect\ion{Fe}{2} fluxes for the model~A with \protect\lya\ and \protect\lyb\ pumping.}
\tablehead{
\colhead{$\lambda$(\AA)} & \colhead{Upper} & \colhead{Lower} & \colhead{$E_{l}$(eV)} & \colhead{Flux} &
\colhead{$\lambda$(\AA)} & \colhead{Upper} & \colhead{Lower} & \colhead{$E_{l}$(eV)} & \colhead{Flux}
}
\startdata
2749.482 & $z\,^{4}D^{o}_{1/2}$ & $a\,^{4}D^{e}_{1/2}$ & 1.096 & $1.61\,^{+ 3}$ & 2858.629 & $d\,^{4}D^{e}_{3/2}$ & $z\,^{4}D^{o}_{1/2}$ & 5.603 & $6.99\,^{+ 3}$ \\
2749.484 & $u\,^{2}F^{o}_{5/2}$ & $c\,^{4}F^{e}_{7/2}$ & 6.220 & $8.87\,^{+ 2}$ & 2861.171 & $z\,^{6}P^{o}_{3/2}$ & $a\,^{4}D^{e}_{3/2}$ & 1.076 & $9.60\,^{+ 2}$ \\
2752.151 & $b\,^{6}D^{e}_{3/2}$ & $z\,^{6}F^{o}_{3/2}$ & 5.255 & $1.50\,^{+ 3}$ & 2864.121 & $b\,^{6}D^{e}_{5/2}$ & $z\,^{6}P^{o}_{3/2}$ & 5.406 & $1.71\,^{+ 3}$ \\
2753.024 & $v\,^{4}D^{o}_{1/2}$ & $c\,^{4}F^{e}_{3/2}$ & 6.206 & $2.46\,^{+ 3}$ & 2865.454 & $d\,^{4}D^{e}_{3/2}$ & $z\,^{4}F^{o}_{3/2}$ & 5.613 & $4.55\,^{+ 3}$ \\
2753.291 & $z\,^{2}I^{o}_{11/2}$ & $b\,^{2}H^{e}_{9/2}$ & 3.266 & $6.06\,^{+ 2}$ & 2868.875 & $z\,^{6}P^{o}_{5/2}$ & $a\,^{4}D^{e}_{5/2}$ & 1.040 & $2.34\,^{+ 3}$ \\
2754.889 & $b\,^{6}D^{e}_{5/2}$ & $z\,^{6}F^{o}_{7/2}$ & 5.235 & $7.35\,^{+ 3}$ & 2869.317 & $d\,^{4}D^{e}_{5/2}$ & $z\,^{4}D^{o}_{3/2}$ & 5.582 & $1.02\,^{+ 4}$ \\
2755.739 & $z\,^{4}F^{o}_{9/2}$ & $a\,^{4}D^{e}_{7/2}$ & 0.986 & $8.92\,^{+ 3}$ & 2872.258 & $d\,^{4}D^{e}_{5/2}$ & $z\,^{4}F^{o}_{5/2}$ & 5.587 & $3.49\,^{+ 3}$ \\
2761.811 & $z\,^{4}D^{o}_{3/2}$ & $a\,^{4}D^{e}_{1/2}$ & 1.096 & $1.97\,^{+ 3}$ & 2874.856 & $z\,^{6}P^{o}_{3/2}$ & $a\,^{4}D^{e}_{1/2}$ & 1.096 & $8.53\,^{+ 2}$ \\
\enddata
\end{deluxetable}

\clearpage
\addtocounter{table}{-1}

\begin{deluxetable}{rllrr||rllrr}
\footnotesize
\tablewidth{18cm}
\tablecaption{\protect\ion{Fe}{2} fluxes for the model~A with \protect\lya\ and \protect\lyb\ pumping.}
\tablehead{
\colhead{$\lambda$(\AA)} & \colhead{Upper} & \colhead{Lower} & \colhead{$E_{l}$(eV)} & \colhead{Flux} &
\colhead{$\lambda$(\AA)} & \colhead{Upper} & \colhead{Lower} & \colhead{$E_{l}$(eV)} & \colhead{Flux}
}
\startdata
3033.437 & $d\,^{4}D^{e}_{1/2}$ & $z\,^{4}P^{o}_{3/2}$ & 5.873 & $1.10\,^{+ 3}$ & 3945.214 & $z\,^{6}D^{o}_{3/2}$ & $a\,^{4}P^{e}_{3/2}$ & 1.695 & $4.96\,^{+ 3}$ \\
3036.967 & $d\,^{4}D^{e}_{5/2}$ & $z\,^{4}P^{o}_{5/2}$ & 5.821 & $4.62\,^{+ 3}$ & 4002.098 & $z\,^{4}P^{o}_{3/2}$ & $b\,^{4}P^{e}_{1/2}$ & 2.777 & $1.00\,^{+ 3}$ \\
3048.981 & $d\,^{4}D^{e}_{3/2}$ & $z\,^{4}P^{o}_{3/2}$ & 5.873 & $5.70\,^{+ 3}$ & 4173.452 & $z\,^{4}D^{o}_{5/2}$ & $b\,^{4}P^{e}_{5/2}$ & 2.582 & $4.08\,^{+ 3}$ \\
3055.352 & $d\,^{4}D^{e}_{1/2}$ & $z\,^{4}P^{o}_{1/2}$ & 5.902 & $4.66\,^{+ 3}$ & 4178.850 & $z\,^{4}F^{o}_{7/2}$ & $b\,^{4}P^{e}_{5/2}$ & 2.582 & $7.65\,^{+ 3}$ \\
3071.122 & $d\,^{4}D^{e}_{3/2}$ & $z\,^{4}P^{o}_{1/2}$ & 5.902 & $3.84\,^{+ 3}$ & 4233.163 & $z\,^{4}D^{o}_{7/2}$ & $b\,^{4}P^{e}_{5/2}$ & 2.582 & $9.99\,^{+ 3}$ \\
3076.432 & $d\,^{4}D^{e}_{5/2}$ & $z\,^{4}P^{o}_{3/2}$ & 5.873 & $7.97\,^{+ 3}$ & 4258.153 & $z\,^{4}F^{o}_{3/2}$ & $b\,^{4}P^{e}_{3/2}$ & 2.703 & $2.13\,^{+ 3}$ \\
3078.678 & $d\,^{4}D^{e}_{7/2}$ & $z\,^{4}P^{o}_{5/2}$ & 5.821 & $1.90\,^{+ 4}$ & 4273.314 & $z\,^{4}D^{o}_{1/2}$ & $b\,^{4}P^{e}_{3/2}$ & 2.703 & $1.08\,^{+ 3}$ \\
3161.976 & $z\,^{4}F^{o}_{3/2}$ & $a\,^{4}P^{e}_{3/2}$ & 1.695 & $2.76\,^{+ 3}$ & 4296.572 & $z\,^{4}F^{o}_{5/2}$ & $b\,^{4}P^{e}_{3/2}$ & 2.703 & $4.49\,^{+ 3}$ \\
\enddata
\end{deluxetable}

\clearpage
\addtocounter{table}{-1}

\begin{deluxetable}{rllrr||rllrr}
\footnotesize
\tablewidth{18cm}
\tablecaption{\protect\ion{Fe}{2} fluxes for the model~A with \protect\lya\ and \protect\lyb\ pumping.}
\tablehead{
\colhead{$\lambda$(\AA)} & \colhead{Upper} & \colhead{Lower} & \colhead{$E_{l}$(eV)} & \colhead{Flux} &
\colhead{$\lambda$(\AA)} & \colhead{Upper} & \colhead{Lower} & \colhead{$E_{l}$(eV)} & \colhead{Flux}
}
\startdata
5316.776 & $z\,^{4}D^{o}_{5/2}$ & $a\,^{4}G^{e}_{7/2}$ & 3.220 & $2.29\,^{+ 3}$ & 9210.938 & $u\,^{4}D^{o}_{1/2}$ & $d\,^{4}D^{e}_{1/2}$ & 9.957 & $1.96\,^{+ 3}$ \\
5325.541 & $z\,^{4}F^{o}_{7/2}$ & $a\,^{4}G^{e}_{7/2}$ & 3.220 & $6.64\,^{+ 3}$ & 9296.851 & $u\,^{4}D^{o}_{5/2}$ & $d\,^{4}D^{e}_{5/2}$ & 9.900 & $2.47\,^{+ 3}$ \\
5362.847 & $z\,^{4}D^{o}_{7/2}$ & $a\,^{4}G^{e}_{9/2}$ & 3.198 & $6.91\,^{+ 3}$ & 9325.115 & $u\,^{4}P^{o}_{3/2}$ & $d\,^{4}D^{e}_{3/2}$ & 9.937 & $8.83\,^{+ 2}$ \\
6147.738 & $z\,^{4}P^{o}_{1/2}$ & $b\,^{4}D^{e}_{3/2}$ & 3.887 & $1.65\,^{+ 3}$ & 9377.042 & $u\,^{4}P^{o}_{1/2}$ & $d\,^{4}D^{e}_{3/2}$ & 9.937 & $1.35\,^{+ 3}$ \\
6149.232 & $z\,^{4}P^{o}_{1/2}$ & $b\,^{4}D^{e}_{1/2}$ & 3.888 & $1.19\,^{+ 3}$ & 9406.669 & $x\,^{6}F^{o}_{9/2}$ & $d\,^{4}D^{e}_{7/2}$ & 9.845 & $2.96\,^{+ 3}$ \\
6238.419 & $z\,^{4}P^{o}_{3/2}$ & $b\,^{4}D^{e}_{3/2}$ & 3.887 & $1.62\,^{+ 3}$ & 9572.625 & $d\,^{4}P^{e}_{5/2}$ & $z\,^{4}P^{o}_{5/2}$ & 5.821 & $2.12\,^{+ 3}$ \\
6247.574 & $z\,^{4}P^{o}_{3/2}$ & $b\,^{4}D^{e}_{5/2}$ & 3.890 & $1.94\,^{+ 3}$ & 9812.027 & $d\,^{4}P^{e}_{1/2}$ & $z\,^{4}P^{o}_{3/2}$ & 5.873 & $8.87\,^{+ 2}$ \\
6317.982 & $c\,^{4}D^{e}_{7/2}$ & $z\,^{4}D^{o}_{7/2}$ & 5.508 & $5.81\,^{+ 2}$ & 9956.255 & $b\,^{4}G^{e}_{9/2}$ & $z\,^{4}F^{o}_{9/2}$ & 5.482 & $8.40\,^{+ 2}$ \\
\enddata
\end{deluxetable}

\clearpage

\end{document}